%% file: OREXP96-04.tex
%
%
\input vanilla.sty
\input tables.tex
\input epsf

\headline={\ifnum\pageno=1\firstheadline\else
\ifodd\pageno\rightheadline \else\leftheadline\fi\fi}
\def\firstheadline{\hfil}
\def\rightheadline{\hfil}
\def\leftheadline{\hfil}
	\footline={\ifnum\pageno=1\firstfootline\else\otherfootline\fi}
\def\firstfootline{\rm\hss\folio\hss}
\def\otherfootline{\hfil}

 1
 1
 1

\font\tenbf=cmbx10
\font\tenrm=cmr10
\font\tenit=cmti10

\font\eightrm=cmr8
\font\eightit=cmti8

\parindent=1.2pc
\magnification=\magstep1
\hsize=6.0truein
\vsize=8.6truein
\nopagenumbers
\def\RefCDF{1}
\def\RefLCWSF{2}
\def\RefLCWSH{3}
\def\RefPOL{4}
\def\RefTIMB{5}
\def\RefONETOP{6}
\def\RefJKT{7}
\def\RefORR{8}
\def\RefFUJII{9}
\def\RefCJSD{10}
\def\RefKF{11}
\def\RefCHEN{12}
\def\RefTHTH{13}
\def\RefPESKIN{14}
\def\RefFADIN{15}
\def\RefHJK{16}
\def\RefFRANKZ{17}
\def\RefEURO{18}
\def\RefPTH{19}
\def\RefPPH{20}
\def\RefSUMMUR{21}
\def\RefWISRD{22}
\def\RefMILLER{23}
\def\RefPHIL{24}
\def\RefSUZUKI{25}
\def\RefFC{26}
\def\RefMPCS{27}
\def\RefYUAN{28}
\def\RefSCHMIDT{29}
\def\RefTAUCHI{30}
\def\RefCP{31}

\rightline{OREXP 96-04}
\bigskip
\centerline{\tenbf TOP QUARK PHYSICS AT A FUTURE $e^+e^-$ COLLIDER:}
\baselineskip=13pt
\centerline{\tenbf EXPERIMENTAL ASPECTS%
\footnote{
{\it Presented at Conference on Physics and Experiments with Linear Colliders,
\hfil\break Morioka-Appi, Japan, September 1995.}\hfil\break
Supported by Department of Energy contract DE-FG06-85ER40224.
}}

\centerline{\eightrm RAYMOND FREY}
\baselineskip=12pt
\centerline{\eightit Physics Department, University of Oregon}
\baselineskip=10pt
\centerline{\eightit Eugene, Oregon 97403, USA}
\centerline{\eightrm E-mail: rayfrey\@bovine.uoregon.edu}\vglue0.2cm

\vglue0.6cm
\centerline{\eightrm ABSTRACT}
\vglue0.2cm
{\rightskip=3pc
 \leftskip=3pc
 \eightrm\baselineskip=10pt\noindent
An overview of top physics and phenomenology at a high-energy linear
collider is presented. A comprehensive study of top quark physics is possible
at such a facility. The unique threshold production of top pairs would provide 
measurements of fundamental properties, such as mass and total decay width,
to unmatched precision. Above threshold, 
the full set of Standard Model and anomalous
electroweak top couplings can be readily measured with
excellent precision. It should also be possible to measure the top Yukawa
coupling. This set of measurements would allow a definitive test of the
widely held notion that the top quark may play a special role in physics
beyond the Standard Model.

\vglue0.6cm}

\tenrm\baselineskip=13pt
\leftline{\tenbf 1. Introduction}
\vglue0.4cm

The stage for the future of top physics has been set by the discovery of top
at Fermilab this year. While the existence of the top quark has long been expected,
its discovery represents a tremendous accomplishment, especially given the
incredibly large value of its mass. The published mass values by the CDF and
D0 collaborations$^{\RefCDF}$ ~are $176\pm 8\pm10$ GeV/c$^2$ and 
$199^{+19}_{-21}\pm 22$ GeV/c$^2$, respectively. Thus, we not only have a mass
value of $\approx 180\pm 12$ GeV/c$^2$ to use for physics studies at future
facilities, but such a large value forces one to consider the distinct possibility
that the top quark plays a special role in particle physics. At the very least,
the properties of the top quark should give important hints of any new physics.

In this context, the determination of a complete set of top properties should be
an important goal of the field. A high-energy future linear $e^+e^-$ collider (FLC)
provides a very
impressive tool to carry out a detailed top-quark physics program. These
capabilities have been reported in the previous workshops in this 
series.$^{\RefLCWSF,\RefLCWSH}$
To a large part, these studies made the top physics case for 
the FLC. The large mass sets a new tone for the studies, as alluded to above,
and in some cases changes the qualitative and quantitative aspects of the
measurements. The $t\bar{t}$ threshold becomes less distinctive, but also less
sensitive to accelerator effects. The measurement of top-quark couplings becomes
an increasingly important overall goal of FLC physics. The Higgs Yukawa coupling
is a fundamental element of the Standard Model, and in non-minimal models of
electroweak symmetry breaking, the Yukawa couplings are modified. But since
this coupling is proportional to mass, the top quark will likely offer the
only possibility for its measurement. There has been much speculation that
the large top mass is an indication that, indeed, top has a direct role in
the physics of electroweak symmetry breaking, and the large top Yukawa coupling
is an indication of this special role.

Nonetheless, there has been some real progress in the FLC top physics analyses 
since the last workshop. As the accelerator designs have progressed, the 
parameters which affect the physics have become better determined. The success
of the polarized electron beam program at SLC$^{\RefPOL}$ implies that a
highly polarized electron beam ($\geq 80\%$) at FLC is 
easily achievable. This is
an important ingredient for the physics, and has been increasingly applied to
the FLC studies. The study of couplings directly benefits from polarized beam,
and all top physics benefits from the dramatic reduction of the $W^+W^-$ final
state with right-handed electron beam. It is also clear that some elements of
experimentation at a FLC have yet to be included. In particular, it is clear that
secondary vertex detection capabilities should be excellent at the FLC, with
important implications for event selection efficiency and purity.

\vglue0.6cm
\leftline{\tenbf 2. Top Production and Decay }
\vglue0.4cm 

The production of top quark pairs in $e^+e^-$ annihilation near threshold is
discussed in more detail in the next chapter. Here, we introduce some basic
features of Standard Model (SM) open top production, SM top decay, 
and some broad implications for detection of top decays. 
There are important radiative effects in high-energy $e^+e^-$
collisions, primarily from initial-state bremsstrahlung and
from beamstrahlung, which arises from the large electromagnetic
fields produced by the tightly focussed beams at the 
interaction point. However, due to the energy dependence of
the top-pair threshold region, this piece of FLC physics is perhaps 
most strongly affected by these radiative phenomena.  These 
effects are introduced in Section 2.2, and are discussed
in more detail in the context of threshold physics in the 
next chapter. Most of the basic production and decay
information in this section exists in more detail in the 
proceedings of the previous meetings$^{\RefLCWSF,\RefLCWSH}$ in this series.

The $t\bar{t}$ cross section due to $s$-channel $e^+e^-$
annihilation mediated by $\gamma,Z$ bosons increases abruptly
at threshold (see Fig. 1), reaches a maximum roughly 50 GeV
above threshold, then falls roughly as the point cross section
($\sigma_{pt}=87({\tenrm fb})/s({\tenrm TeV})$). At $\sqrt{s}=500$
GeV the lowest-order total cross section for unpolarized beams
is $0.54$ pb; it is $0.74$ ($0.34$) for a fully left-hand 
(right-hand) polarized electron beam. With increasing energy
$t$-channel processes resulting, for example, in final states such
as $e^+e^-t\bar{t}$ or $\nu\bar{\nu}t\bar{t}$, have increasing
cross sections. However, even at 1 TeV these cross sections are still 
much smaller than those due to the annihilation process, and
they are not considered further here.
However, in the context of tests of strongly-coupled electroweak
symmetry breaking, this $\nu\bar{\nu}t\bar{t}$ process may be of 
particlular interest, as suggested$^{\RefTIMB}$ at this meeting.
Theoretical results on
single-top production via the process $e^+e^-\rightarrow e\nu t b$
were also presented here,$^{\RefONETOP}$ and given very
high luminosity running at high energy, this offers some
attractive physics possibilities, particularly for a $V_{tb}$ 
measurement. Processes involving Higgs
production or exchange, while also having relatively small cross
sections, offer the exciting possibility of measuring the Higgs
Yukawa coupling. This issue is considered in Chapters 2 and 5.

By far the dominant new influence on top phenomenology is a
direct consequence of its large mass: The very large decay 
width. In the Standard Model the weak decay of top proceeds
very rapidly via $t\rightarrow bW$ according to
$$ \Gamma_t = 0.18 (m_t/m_W)^3 \eqno{(1)} $$
For $m_t=180$ GeV/c$^2$ this lowest-order prediction is 
$\Gamma_t = 1.71$ GeV.  After first-order QCD and electroweak
corrections,$^{\RefJKT}$ this becomes $1.57$ GeV.
Hence, top decay is much more rapid than the 
characteristic time for hadron formation, for which the
scale is $\Lambda_{\tenrm QCD}^{-1}$. 

This implies that the 
phenomenology of top physics is fundamentally 
different than that of the lighter quarks. First of all,
there will be no top-flavored mesons. While we lose the familiar
study of the spectroscopy of these states, we gain unique clarity
in the ability to reconstruct the final state. This may prove to
be a crucial advantage toward uncovering fundamental issues.
The top decay also provides a natural cutoff for gluon emission.
In fact, the color strings form along the separating $b$ and 
$\bar{b}$ quarks. The character of the interference between gluons 
emitted from top and bottom quarks therefore depends upon
the value of $\Gamma_t$.$^\RefORR$

The parton-level decay of top implies that, unlike other quarks, 
the top spin is transferred to a readily reconstructable final state. 
Measurement of the $b\bar{b}W^+W^-$ final state therefore
provides a powerful means of probing new physics manifested
by top with helicity analyses. This is explored in Section 4.2.
Another interesting implication of the large $m_t$ is the
SM prediction that the decay $t\rightarrow bW$ produces mostly
longitudinally polarized W bosons, with a degree of longitudinal
polarization given by $m_t^2/(m_t^2 + 2M_W^2)\approx 72\%$ for  
$m_t=180$ GeV/c$^2$. This is, in itself, an interesting fact
given the important role of longitudinally polarized W bosons
in electroweak symmetry breaking.
  
The luminosity parameters vary somewhat between the
various FLC designs. A typical luminosity at $\sqrt{s}=500$ GeV
is $5\times 10^{33}$ cm$^{-2}$s$^{-1}$, and increases by about
a factor two at $\sqrt{s}=1$ TeV, while the annihilation cross 
section drops by a factor four. Hence, for a typical assumption
of $10^7$ seconds of useful running per year, a design year of
integrated luminosity at $\sqrt{s}=500$ GeV is 50 fb$^{-1}$.
This corresponds to roughly $25\times 10^3$ produced $t\bar{t}$
events per design year. An important advantage of $e^+e^-$ 
colliders is, of course, that the majority of produced events can
be used in the physics analyses. Nevertheless, one can anticipate
that statistical errors will dominate many measurements. In fact,
one of the primary goals of the physics studies is to estimate
whether the dominant systematic errors are indeed small.

In the Standard Model $| V_{tb} |\approx 1$, so that the
decay mode $t\rightarrow bW$ completely saturates the decay
width of Eq.~1. Ignoring hard-gluon radiation, the final state
is given by the W decay modes from the $b\bar{b}W^+W^-$
intermediate state. Hence, we have the following 
lowest-order (corrected) decay fractions:
BR$(t\bar{t}\rightarrow b\bar{b}q q^\prime qq^\prime)
= 36/81\;(0.455)$;  BR$(t\bar{t}\rightarrow b\bar{b} qq^\prime \ell\nu)
= 36/81\;(0.439)$;  BR$(t\bar{t}\rightarrow \ell\nu\ell\nu) 
= 9/81\;(0.106)$,
where $q=u,c$, $q^\prime = d,s$, and $\ell = e,\mu,\tau$. The
numbers in parentheses represent these fractions after QCD
corrections, which produces a factor of $\approx 1.04$ for
$W\rightarrow qq^\prime$ relative to $W\rightarrow\ell\nu$.

Non-standard top decays would, of course, be an interesting
addition to top physics. The measurement of the top decay width
gives an indirect indication of appreciable new decay modes, and
is measurable from the threshold at the level of $10\%$ or better.
Studies of a few non-standard decays have been studied. In
particular, studies of the modes $t\rightarrow bH^+$ and 
$t\rightarrow \tilde t\tilde\chi$ have been presented$^\RefLCWSH$
and shown to be readily separated with straightforward cuts.
It would also be interesting to study the capability for a
direct search for non-$b$ top decays. For example, there has been
much theoretical interest in $t$-$c$-Higgs couplings which could
give rise to $e^+e^-\rightarrow t\bar{c}$, $\bar{t}c$.

\vglue0.6cm
\leftline{\tenit 2.1. Event Measurement }
\vglue0.4cm 

Event selection and backgrounds, as discussed in previous 
reports,$^{\RefLCWSF,\RefLCWSH}$ is briefly summarized here. The $t\bar{t}$
cross section at $\sqrt{s}=500$ GeV is roughly $0.5$ pb. On the other
hand, the cross section for lepton and light quark pairs
is about 16 pb, while for $W^+W^-$ production it is about 8 pb.
The emphasis of most event selection strategies has been to take advantage
of the multi-jet topology of the roughly $90\%$ of $t\bar{t}$
events with 4 or 6 jets in the final state. Therefore, cuts on
thrust or number of jets drastically reduces the light fermion
pair background. In addition, one can use the multi-jet mass
constraints $M($jet-jet$)\approx M_W$ and
$M($3-jet$)\approx m_t$ for the cases involving
$t\rightarrow bqq^\prime$. Simulation studies$^{\RefFUJII}$
have shown that multi-jet resolutions of 5 GeV/c$^2$ and
15 GeV/c$^2$ for the 2-jet and 3-jet masses, respectively,
are adequate and readily achievable with standard resolutions.
A detection efficiency of about 70\% 
with a signal to background ratio of 10 was attained in selecting
6-jet final states just above threshold. These numbers are typical
also for studies which select the 4-jet$+\ell\nu$ decay mode.

There are two aspects of event selection which should be
powerful tools, but have not been widely studied. The background
due to W-pair production is the most difficult to eliminate. 
However, in the limit that the electron beam is fully right-hand
polarized, the $W^+W^-$ cross section becomes very small. Hence, even
though the beam polarization will be somewhat less than
 $100\%$, this allows for experimental control and
measurement of the background. On the other hand, the signal is
also reduced, to a much smaller degree, by running with
right-polarized beam. A possible strategy might be
to run with right-hand polarized beam only long enough to make
a significant check of the component of background due to
W pairs. The left-right W-pair asymmetry predictable at the
required few \% level.

Another important technique is that of precision vertex
detection. The working assumption is that the present experience
with SLC/SLD can be used as a model, a rather close model, of
what can be done at a FLC. The small and stable interaction point
of linear $e^+e^-$ colliders, along with the small beam sizes
and bunch-structure timing, make them ideal for pushing the
techniques of vertex detection. The present spatial resolution of
the CCD vertex detector of SLD is 
$9\oplus 29/p({\tenrm GeV})\sin^{3/2}\theta$ $\mu$m in the plane
transverse to the beam ($\theta$ is the angle with
respect to the beamline) and 
$14\oplus 29/p({\tenrm GeV})\sin^{3/2}\theta$ $\mu$m in the $r-z$
plane. Equally important is that the primary interaction point is
determined equally well. The upshot is that single-hemisphere
$b$-tagging is now done with $\sim 40\%$ efficiency and $99\%$ purity,
and should improve at a FLC. 

Clearly this should have a big impact on FLC top physics, where
every event has two high-momentum bottom jets. An event
selection strategy should include loose $b$-tagging criteria. 
For example, a high efficiency for tagging one $b$ or the other
($\sim 90\%$) could be achieved with reduced purity. This, 
combined with loose topological and mass cuts to reduce background
should provide a very efficient and pure selection. 
A possible scenario for FLC vertex detection was 
presented$^{\RefCJSD}$ at this meeting. With a large solenoidal
magnetic field ($>3$ T), the backgound pairs from the
beam-beam effects would be confined to a small radius, allowing
the inner radius of the CCD detectors to be reduced from 
3 cm with SLD to nearly 1 cm. The relatively short bunch trains
of the SLAC or JLC designs would make CCDs a good technology
choice. The longer TESLA bunch trains may require devices which
have a fast-clear capability. In all cases, the beam backgrounds
increase rapidly at small radius, hence necessitating the use of
some type of pixel technology.

Other detector requirements imposed by top physics are not
particularly special. The one exception is that the measurement
of the luminosity function near threshold may require quite good
spatial resolution for Bhabha-scattered electrons in the
$\sim 200$ mrad (endcap) region of the detector, presumably
involving highly segmented electromagnetic calorimetry. This is
discussed further in Section 3. The masks to absorb the 
beam-induced backgrounds will extend to angles of 100--200
mrad from the beamline. This has a small impact ($\sim 2\%$) on 
the acceptance for top events. One key point is whether the
detector will allow the reconstruction of 6-jet and 8-jet (see
Section 5) final states. Generally, jet reconstruction and 
multi-jet mass resolution is dominated by QCD effects for
modern detectors. However, it is important to see if this is still
the case for such complicated events. 

\vglue0.6cm
\leftline{\tenit 2.2. Radiative and Beam Effects }
\vglue0.4cm 

All FLC physics analyses must consider the effects of
initial-state radiation (ISR) and beamstrahlung (BS) on the 
spectrum of collision energies. Studies of these effects, as well
as the single-beam accelerator energy spread, are presented in
the following sections. The effects of ISR are appreciable for
high energy electron colliders, where the effective expansion
parameter for real photon emission, rather than $\alpha/\pi$, is 
$\beta = {2\alpha\over\pi}(\ln(s/m_e^2) -1) \approx 1/8$
for $\sqrt{s}=500$ GeV. Typically, one can use a calculation
like that of Kuraev and
Fadin$^{\RefKF}$, which sums the real soft-photon emission to
all orders and calculates the initial state
virtual corrections to second order.
For beamstrahlung, the calculation of Chen$^{\RefCHEN}$ provides
a good approximation for the effects of beamstrahlung for
most FLC designs. The figure of merit for the calculation of
beamstrahlung is $\Upsilon = \gamma (B/B_c)$, where 
$\gamma = E_{\tenrm beam}/m_ec^2$, $B$ is the effective magnetic 
field strength of the beam, and 
$B_c = m_e^2c^3/e\hbar \approx 4\times 10^9$ T. 
When $\Upsilon \ll 1$ the beamstrahlung is in the classical
regime and is readily calculated analytically. 
For example, in the case of the SLAC X-band NLC design, we have
$\Upsilon \approx 0.08$ at $\sqrt{s}=500$ GeV. In this case, there
is an appreciable probability for a beam electron (or positron) to
emit no photons. So the spectrum is well-approximated as
a delta function at $E=E_{\tenrm beam}$ with a bremsstrahlung-like
spectrum extending to lower energies. As we shall see, the delta
function piece of the spectrum plays an important role in the
shape of the threshold cross section. The NLC design at 500 GeV
yields $43\% $ of the luminosity for which the colliding $e^+e^-$
are unaffected by beamstrahlung. It is interesting to note that
even in this case there are on average $0.91$ emitted photons per
beam electron, so multi-photon emission is an important aspect 
of the calculation.

\vfill\eject

\vglue0.6cm
\leftline{\tenbf 3. Threshold Physics}
\vglue0.4cm

The theoretical underpinnings for the process 
$e^+e^-\rightarrow t\bar{t}$ at threshold have been extensively 
studied,$^{\RefTHTH}$ and will not be
reproduced here. We will briefly discuss the main features of the 
expected threshold physics and the measureable parameters. The 
phenomenolgy associated with the $t\bar{t}$ resonance 
(toponium) is introduced. A number of studies examining 
experimental sensitivity to threshold physics have been 
undertaken. Some of the main implications of these 
studies are presented.

\vglue0.6cm
\leftline{\tenit 3.1. Cross Section  }
\vglue0.4cm

\par\leftskip 0in\rightskip 0in
In Fig.~1 we show the cross section for $t\bar{t}$ 
production as a function of nominal center-of-mass energy for 
$m_t = 180$ GeV/c$^2$. The theoretical cross section, 
indicated as curve (a), is based on the results of 
Peskin and Strassler$^{\RefPESKIN}$ with $\alpha_s(M_Z^2)=0.12$,
infinite Higgs mass, and nominal Standard Model couplings.
Each energy-smearing mechanism, initial-state radiation (b), 
beamstrahlung (c), and beam energy spread (d),  has been 
successively applied. Hence, curve (d) includes all effects. The
beam effects were calculated assuming SLAC X-band NLC design
parameters as an example. In this section, the phenomena
associated with these curves is discussed, along with the role of 
accelerator design for the beam effects.

The threshold enhancement given by the predicted cross section
curve (a) of Fig.~1 reflects the Coulomb-like attraction of the 
produced $t$-$\bar{t}$ due to the short-distance QCD potential
$$ V(r)\sim -C_F{\alpha_s(\mu)\over r} \,,\eqno{(2)}$$
where $C_F=4/3$ and 
$\mu$ is evaluated roughly at the scale of the Bohr radius of this 
$t$-$\bar{t}$ toponium atom (labelled $\theta$):  
$\mu\sim 1/a_\theta = \alpha_s m_t$. This bound state exists,
on average, for approximately one classical revolution before
one of the top quarks undergoes weak decay. The level spacings of the
QCD potential given approximately by the Rydberg energy,
$ \sim\alpha_s^2 m_t$, turn out to be comparable to the 
widths of the resonance states, given by 
$\Gamma_\theta\approx 2\Gamma_t$. Therefore the various
toponium states become smeared together, as seen in Fig.~1,
where only the bump at the position of the 1S resonance is
distinguishable. In fact, for much smaller $m_t$, or for
an anomalously small $\Gamma_t$, the various
states such as 1S, 2P, 2S, etc. are clearly separated in the
theoretical cross section. The infrared cutoff imposed by the
large top width also implies$^{\RefFADIN}$ that the physics is
independent of the long-distance behavior of the QCD potential.
The assumed intermediate-distance potential is also 
found$^{\RefFUJII}$ to have a negligible impact. Hence, the
threshold physics measurements depend on the short-distance
potential of Eq.~2 based on perturbative QCD. At minimum,
a significant qualitative test
of QCD is therefore possible, one which goes beyond simply extracting
the parameter $\alpha_s(M_Z)$. Perhaps it is possible to perform
precise calculations at top threshold, similar to those
performed within the context of lattice QCD for the 
$J/\psi$ or $\Upsilon$ systems, which could be directly
confronted with FLC experiment. 

\medskip
\epsfysize=3.5in \epsfbox[0 0 791 612]{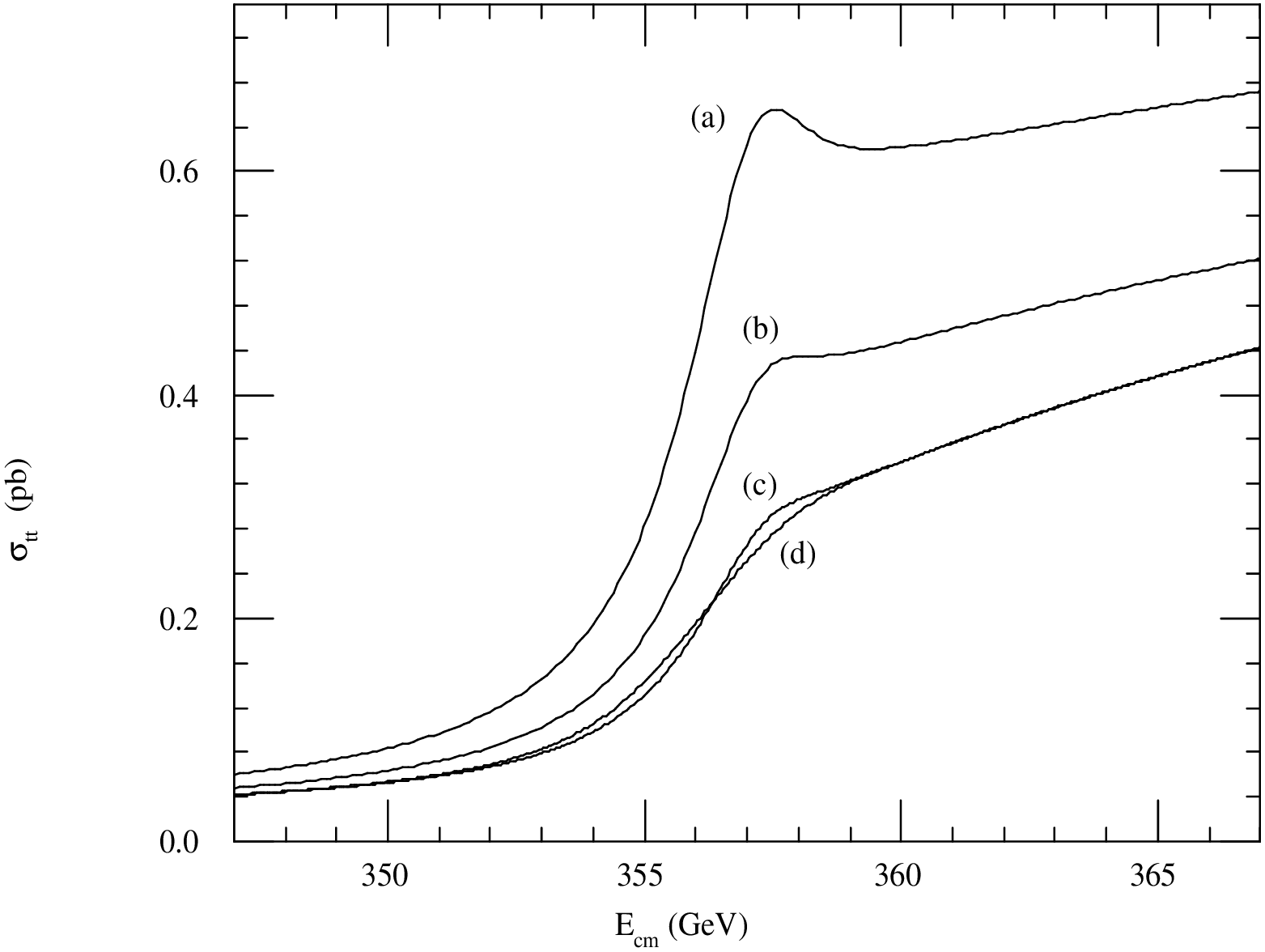}
{\par\noindent\eightrm
Fig.~1. Production cross section for top-quark pairs near 
threshold for $m_t = 180$ GeV/c$^2$. 
The theoretical cross section is given by curve (a), to 
which the energy re-distribution effects have been applied. 
Curve (b): initial-state radiation (ISR); curve (c): ISR and 
beamstrahlung; curve (d): ISR, beamstrahlung, and 
beam energy spread.}

\bigskip
In increase of $\alpha_s$ deepens the QCD potential, thereby
increasing the wave function at the origin and producing an
increased 1S resonance bump. In addition, the binding energy of
the state increases roughly as the Rydberg energy 
$\sim \alpha_s^2 m_t$. So the larger $\alpha_s$ has the
combined effect of increasing the cross section as well as
shifting the curve to lower energy. The latter effect is also
what is expected for a shift to lower $m_t$. Therefore, there
exists a significant correlation between the measurments of
$\alpha_s$ and $m_t$ from a threshold scan. This is evident
from Fig.~10b.

The total top decay width, $\Gamma_t$, is an essential piece of
exploratory top physics. It is intrinsic to the threshold
shape, and is perhaps best measured at threshold. 
For a quarkonium state, we expect the cross section at the
1S peak to vary with the total width roughly as 
$\sigma_{1S}\sim |V_{tb}|/\Gamma_t $, and therefore is very
sensitive to the width, as indicated by Fig.~2 for
rather wide variations in $\Gamma_t$ relative to the Standard
Model expectation. It is noted that the calculations of Figs.~1
and 2 use the uncorrected top width, as discussed in Section 2.1,
so that the resonance structure will actually be more 
distinctive after correction than that which is shown
in these figures.  Combining
the cross section information with the momentum and asymmetry
results, as discussed below, represents what is most likely the
best opportunity to measure $\Gamma_t$.

\epsfysize=3.0in \epsfbox[0 0 791 612]{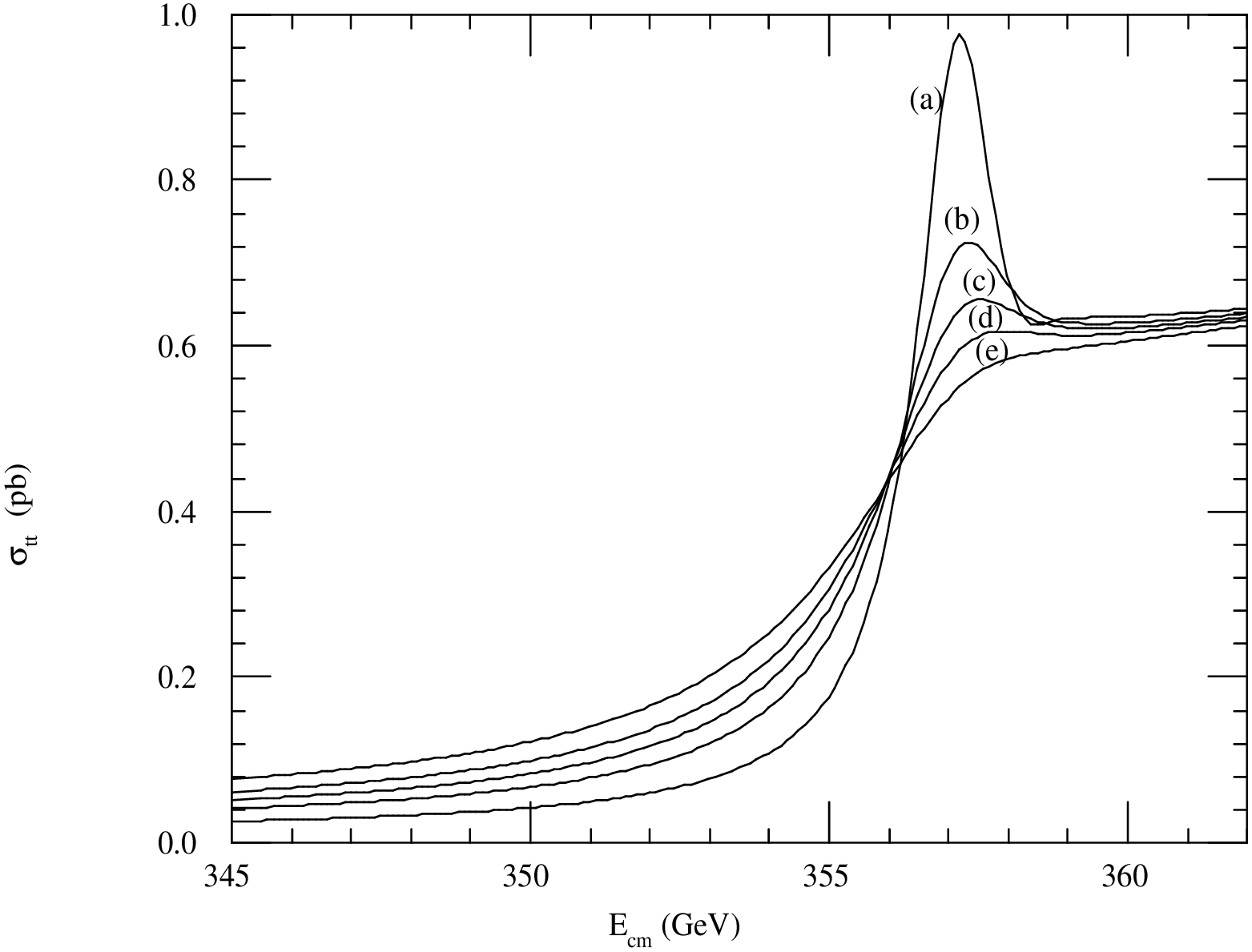}
{\par\noindent\leftskip 0.5in\rightskip 0.5in\eightrm
Fig.~2. Variation of theoretical threshold cross section with top width
for $m_t=180$ GeV/c$^2$. The curves correspond to values of 
$\Gamma_t/\Gamma_{SM}$ of (a) 0.5; (b) 0.8; (c) 1.0; (d) 1.2; (e) 1.5.}

\bigskip
In addition to the QCD potential, the $t$-$\bar{t}$ pair is also
subject to the Yukawa potential associated with Higgs exchange:
$$ V_Y = -{{\lambda^2}\over{4\pi}} {{e^{-m_H r}}\over{r}} 
\,,\eqno{(3)}$$
where $m_H$ is the Higgs mass and $\lambda$ is the Yukawa
coupling, which in the Standard Model is
$$\lambda = \bigl[\sqrt{2}G_F\bigr]^{1/2}\,m_t \eqno{(4)} $$
Because of the extremely short range of the Yukawa potential,
its effect is primarily to alter the wave function at the origin,
and hence to shift the level of the cross section across the
1S resonance. The Higgs effect at threshold has been carefully
calculated$^{\RefHJK}$ with results shown in Fig.~3. Presumably, the
Higgs boson responsible for this enhancement will have already been
discovered. So the importance of this measurement would be to
check the SM relationship of Eq.~4, which was used in the calculation
of Fig.~3. 

\bigskip
\epsfysize=3in \epsfbox{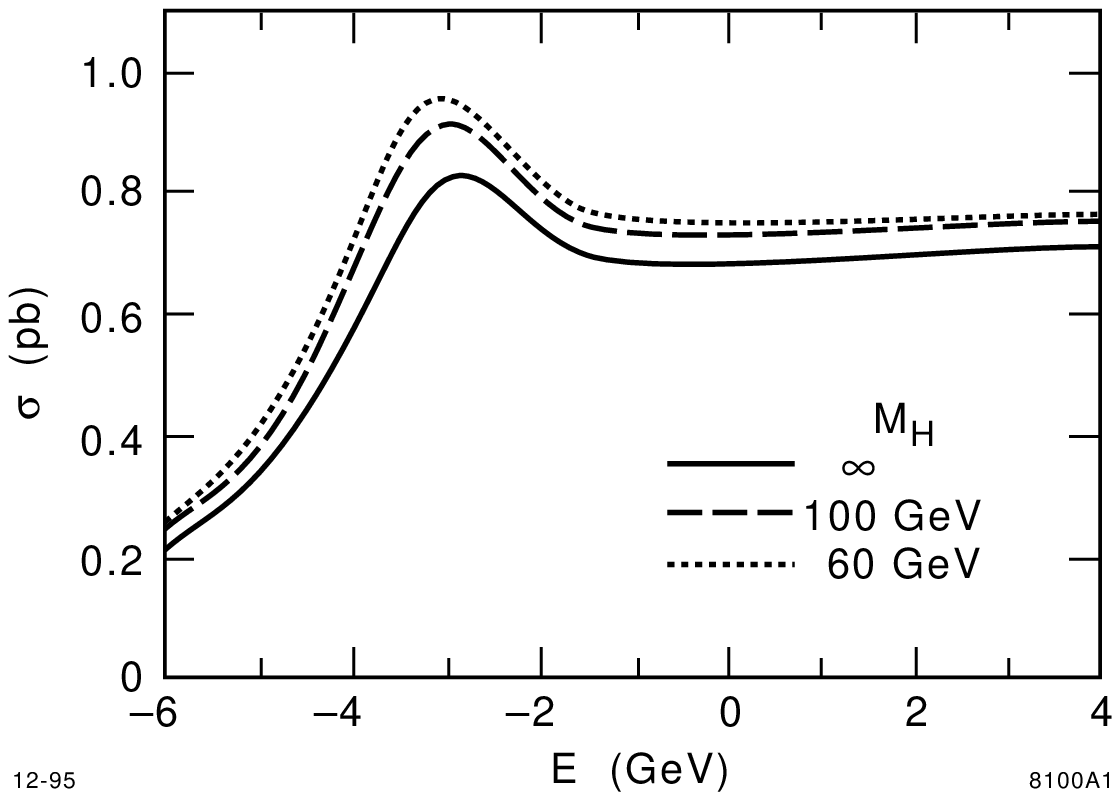}
{\par\noindent\eightrm
\centerline{Fig.~3. Theoretical cross section as a function of
Higgs mass, from Ref.~\RefHJK.}}

\bigskip
This is an exciting possibility, and underscores what could be
gained by performing careful threshold measurements.
The physics of the threshold cross section is, in 
summary, expected to depend on the following set of parameters:
$$\sigma = \sigma (m_t,\alpha_s,\Gamma_t,m_H,\lambda)
\eqno{(5)} $$
The experimental challenge is to unravel the various dependencies,
each of which is qualitatively different, using the cross section scan,
as well as the momentum and forward-backward asymmetry measurements
discussed in Section 3.3. To do so requires understanding the
center-of-mass energy loss mechanisms discussed below.

\vfill\eject

\vglue0.6cm
\leftline{\tenit 3.2. Radiative and Beam Effects  }
\vglue0.4cm

The general framework in which calculations of radiative effects
was introduced in Section 2.2.  Because of the relatively narrow
energy structure associated with the top threshold, these effects
play a special role. Qualitatively, we can understand these effects
in the following way. Both ISR and beamstrahlung (BS) spectra 
are peaked at zero energy loss, and have long, 
rather low tails extending to large energy loss. As mentioned
previously, the BS spectra is, in fact, quite well modelled by a
combination of delta function at zero energy loss plus a
synchrotron-like tail. The delta function part, of course, does
nothing to disturb the resolution of the threshold. 
Once the energy loss is greater than $\sim\Gamma_t$, then the
corresponding luminosity is lost for threshold physics, but 
it does not contribute to a smearing in energy of the threshold
structure.  It is only those portions of the ISR or BS spectra 
with energy loss $<\Gamma_t$ which contribute to the
smearing effects. 

Curves (b) and (c) of Fig.~1 indicate the effects of ISR and 
ISR combined with BS, respectively. Initial-state radiation,
of course depends only on the center-of-mass energy. On the
other hand, the accelerator designers have some control 
of the beamstrahlung, which depends on a number of parameters,
but can be grossly characterized by the mean field strength
$ B = 6\times 10^2$ T. The beamstrahlung calculation of Fig.~1
assumes an NLC X-band design with a parameter set optimized
for $\sqrt{s}=500$ GeV, with spot sizes $\sigma_x$ and $\sigma_y$
simply scaled to 360 GeV and with no change in $\beta$
functions. This is clearly not optimal for luminosity, but should 
gives a reasonable estimate for the beamstrahlung. 

An additional accelerator effect on the threshold shape
is that which results from the energy spread of each individual
beam in its respective LINAC. The energy spread for a single beam,
$\Delta E/E$, can also, to some extent, be controlled by both
design and tuning. Generally, a tunable range of $\pm 50$ is possible.
Typically, a reduced energy spread results in lower luminosity, while
a larger spread produces more backgrounds. As discussed below, the
distributions of $\Delta E/E$ at a linear collider,  vary according to
details of energy compression, RF phasing, and so forth, but are
typically not centrally peaked, and so the width is often 
characterized by a FWHM value. This can have a significant effect
on the top threshold shape if the spread is too large. For
$m_t=150$ GeV/c$^2$, the threshold structure is sharper, meaning
that the energy spread effects are more important (see Ref. 
~\RefFUJII). Curve (d) of Fig.~1 was calculated for a $\Delta E/E$ 
distribution with FWHM of $0.6\%$, combined with ISR and BS
effects. 

\bigskip
\centerline{
\epsfxsize=3in \epsfbox{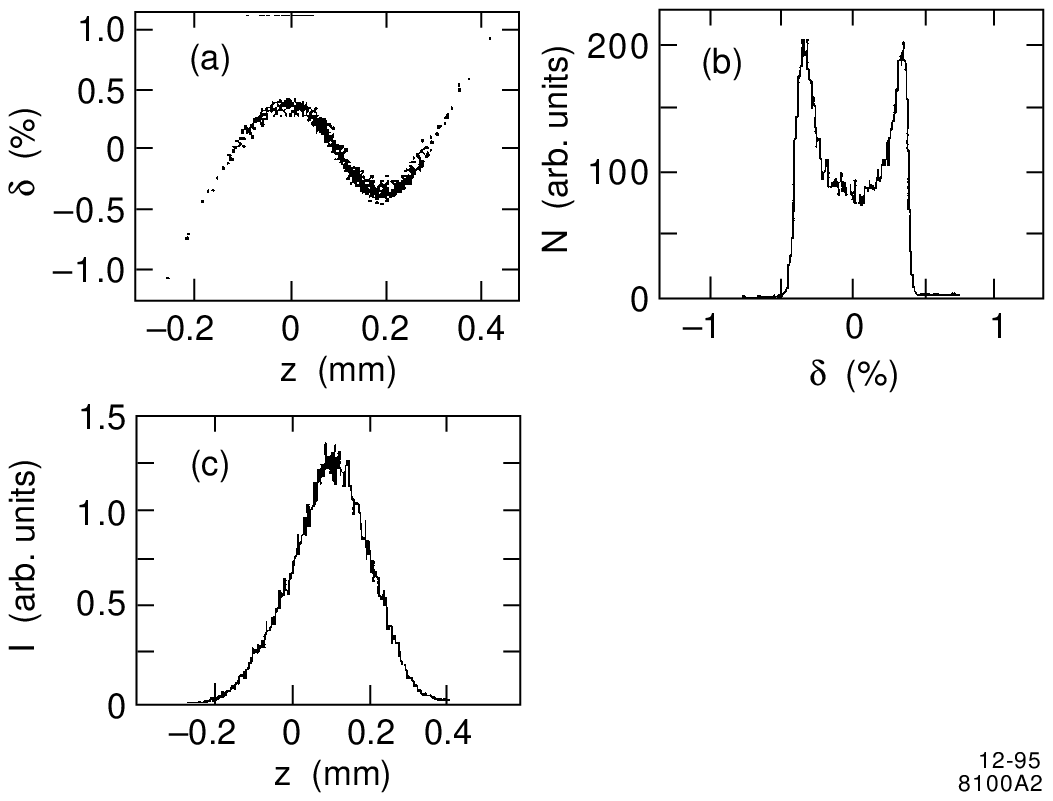}
}
{\par\noindent\leftskip 0.5in\rightskip 0.5in\eightrm
Fig.~4. Expected single-beam energy spread based on SLAC NLC 
design. (a) Scatter plot of single-beam energy spread, 
$\delta\equiv\Delta E/E$, versus longitudinal position within a 
bunch, $z$, along with projections onto
the $\delta$ axis (b), and onto the $z$ axis (c).}

\bigskip
\centerline{
\epsfysize=2.8in \epsfbox{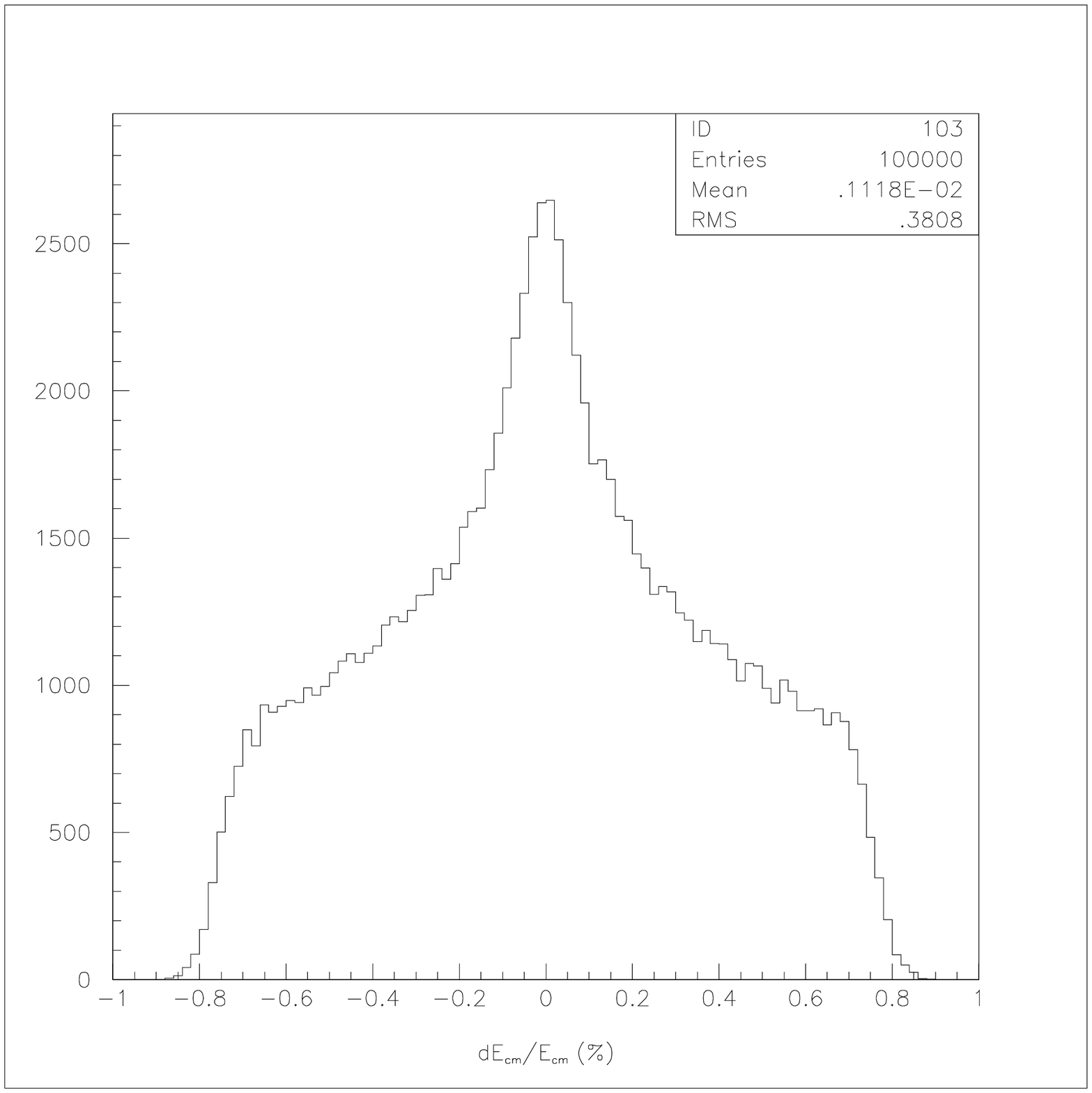}
}
{\par\noindent\leftskip 0.5in\rightskip 0.5in\eightrm
Fig.~5. Distribution of the center-of-mass energy, 
$\Delta E_{cm}/E_{cm}$, due to the convolution of 
single-beam energy spread distributions for the two beams.
The single-beam energy spread in this case has FWHM of $0.8\%$, 
corresponding to the distribution given in Fig.~4.}

\bigskip
A few further comments on the beam energy spread are in order.
Figure 4 depicts the single-beam energy spread 
expected$^{\RefFRANKZ}$ for one possible NLC X-band LINAC 
design. Of course, to determine the effect of these energy
spreads on the threshold, the electron and positron distributions
must be convoluted. This, in principle, can be quite complicated.
However, a reasonable approach is to ignore transverse chromatic
and angular divergence effects, and to simply integrate over
collision length. In this way,
the luminosity-weighted center-of-mass energy spread,
$\Delta E_{cm}/E_{cm}$, is calculated
from the single-beam distribution given its dependence on the 
bunch spatial distribution, as shown in Fig.~4. The resulting 
distribution for $\Delta E_{cm}/E_{cm}$ is given in Fig.~5. 
In this case, the input single-beam 
energy spread has FWHM of $0.8\%$, corresponding to the 
distribution shown in Fig.~4. This distribution, in contrast to the
single-beam case, is strongly peaked at zero with an RMS 
of $0.38\%$, as indicated in the figure. 
For the NLC, it is expected that the
single-beam energy spread can be comfortably adjusted 
within the FWHM interval $0.6\%$ to $1.0\%$, and
for the top threshold, it is clear that the smaller width is 
preferred. Curve (d) of Fig.~1 was calculated using the 
$0.6\%$ width, with a corresponding $\Delta E_{cm}/E_{cm}$
of $0.29\%$. Figure 6 shows the change in shape of the threshold 
cross section as the single-beam energy spread is increased. 
The large top mass of
about 180 GeV/c$^2$ presents a relatively broad, featureless 
threshold shape which is not nearly as sensitive to the 
$\Delta E_{cm}/E_{cm}$ distribution as would
be expected if the top mass were smaller. Since it is not
clear how well this effect could be controlled, this lack of
sensitivity is one welcome outcome of the broader threshold.


\epsfysize=3.1in \epsfbox[0 0 791 612]{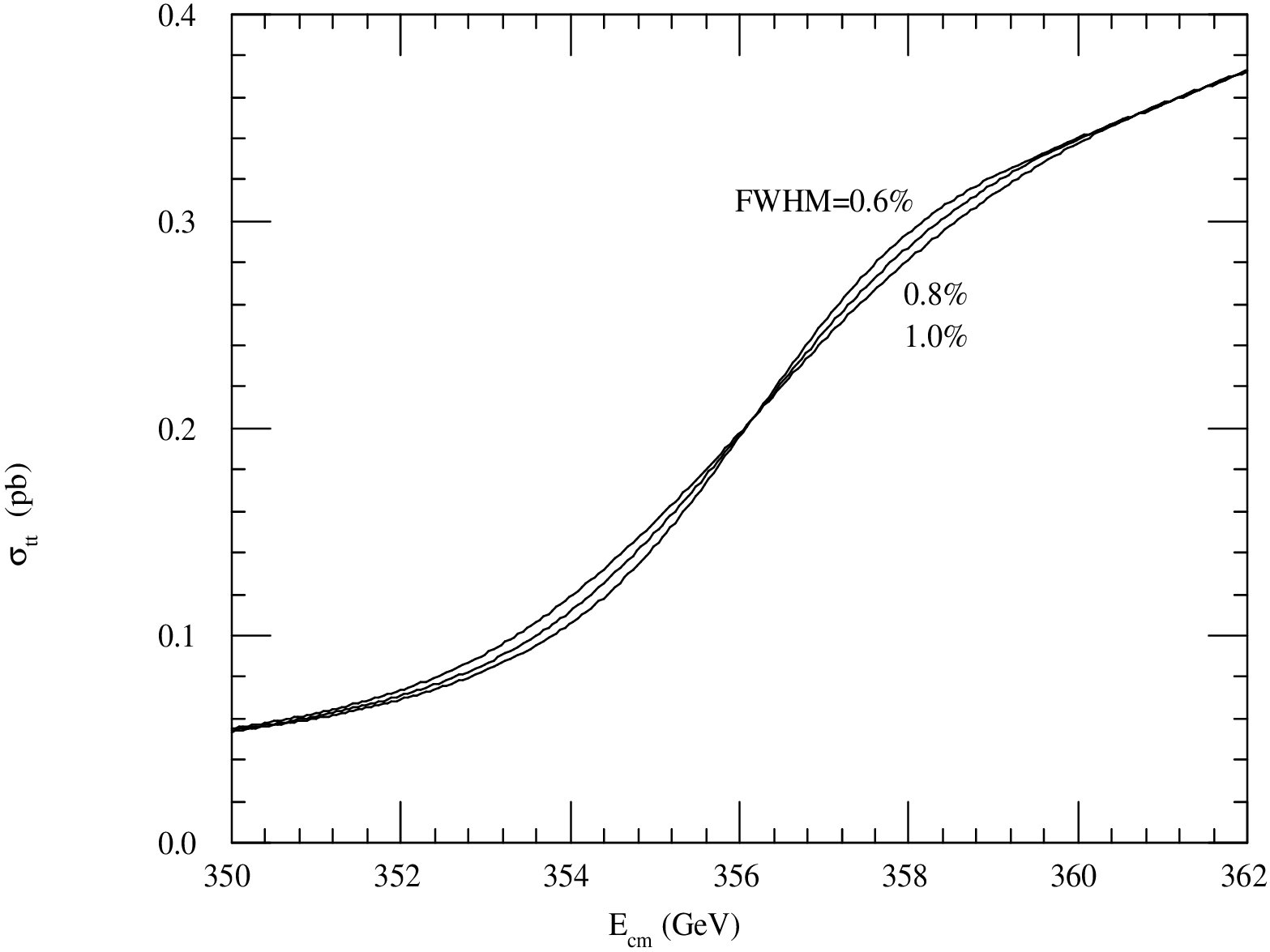}
{\par\noindent\leftskip 0.5in\rightskip 0.5in\eightrm
Fig.~6. Comparison of $t\bar{t}$ threshold shape, with all 
effects included, for different single-beam energy spreads. 
The three curves correspond to single-beam energy 
spread distributions with FWHM of $0.6\%$, $0.8\%$, 
and $1.0\%$, as indicated.}

\bigskip
Since the beamstrahlung and LINAC energy spread effects 
are dependent upon accelerator design, it is interesting
to make comparisons at top threshold, where their
impact is most apparent. Figure 7 represents a calculation
of the top threshold presented$^{\RefEURO}$ at this meeting
for a design based on TESLA.  We see that the accelerator
effects for TESLA are more pronounced than those presented 
in Fig.~1 for the NLC.  This is ascribed to the delta-function
component of the beamstrahlung spectrum, discussed earlier, 
being smaller for the TESLA case. So while the tail of the
BS spectrum for TESLA is not as long as that for NLC, 
which may be advantageous for some physics, the
smaller probability for emitting essentially zero BS energy 
at TESLA results in more energy smearing at top threshold.
The bunch length for TESLA is relatively large, thus enabling
a passing beam particle many opportunities to emit a small,
but non-negligible, amount of radiation.

\epsfysize=3.0in \epsfbox[-100 30 467 567]{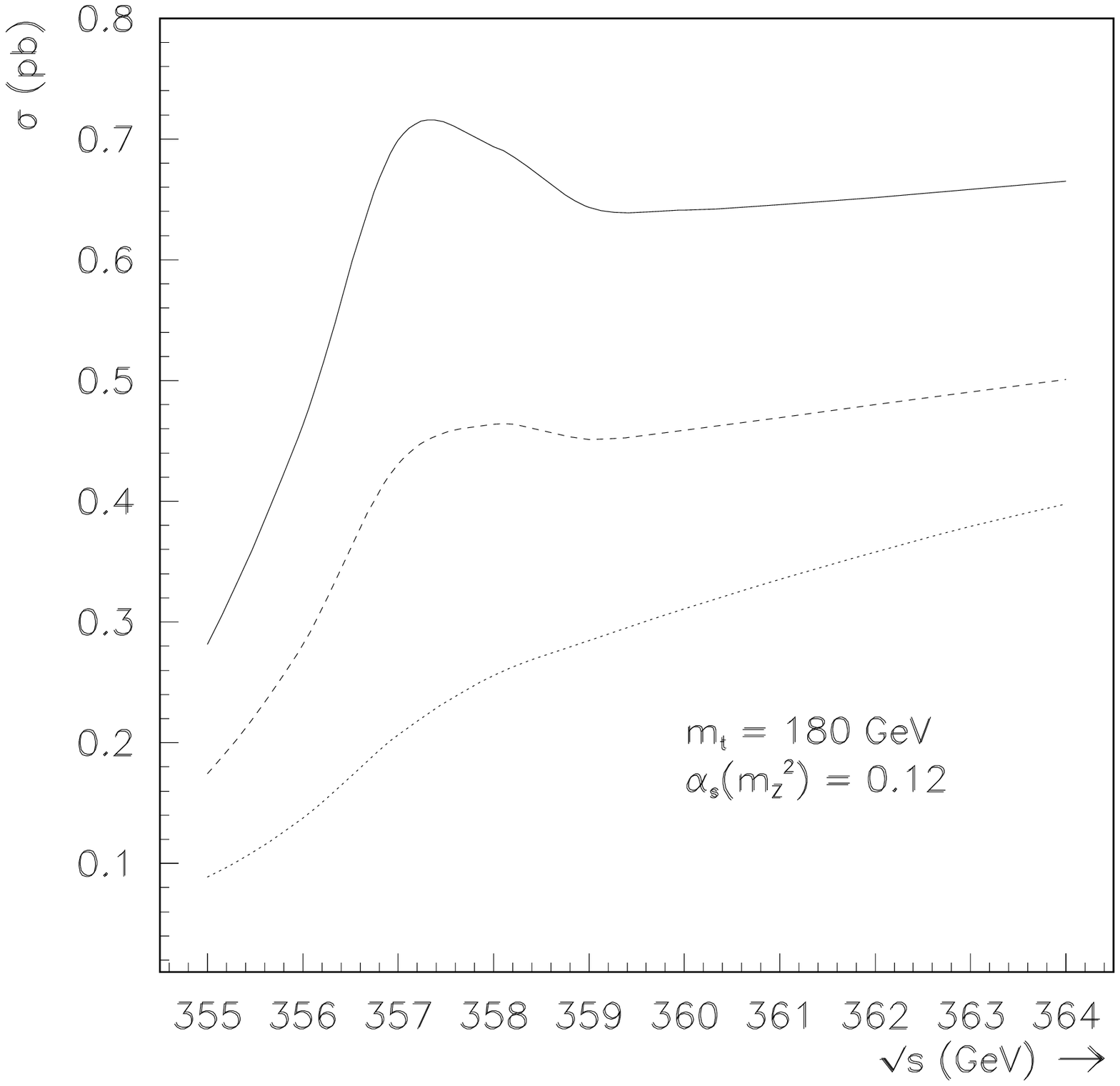}
{\par\noindent\leftskip 0.5in\rightskip 0.5in\eightrm
Fig.~7. Top threshold from Ref.~{\RefEURO}
assuming $m_t=180$ GeV/c$^2$. The accelerator 
effects are calculated for the TESLA design.}

\bigskip

The energy re-distribution effects have an interesting impact
on the dependence of the threshold shape on $\Gamma_t$.
This is shown below in Fig.~8 for which the curves correspond to
the theoretical curves of Fig.~2, but with all effects included, 
in this case assuming NLC parameters. The larger widths give rise
to a much flatter shape, but for which the level below the
1S peak is now altered significantly, and provides the best
discrimination for the measurement. A physics strategy which is
focussed on width optimization would presumably expend a large
fraction of the scan luminosity in this region. 

\epsfysize=3.0in \epsfbox[0 0 791 612]{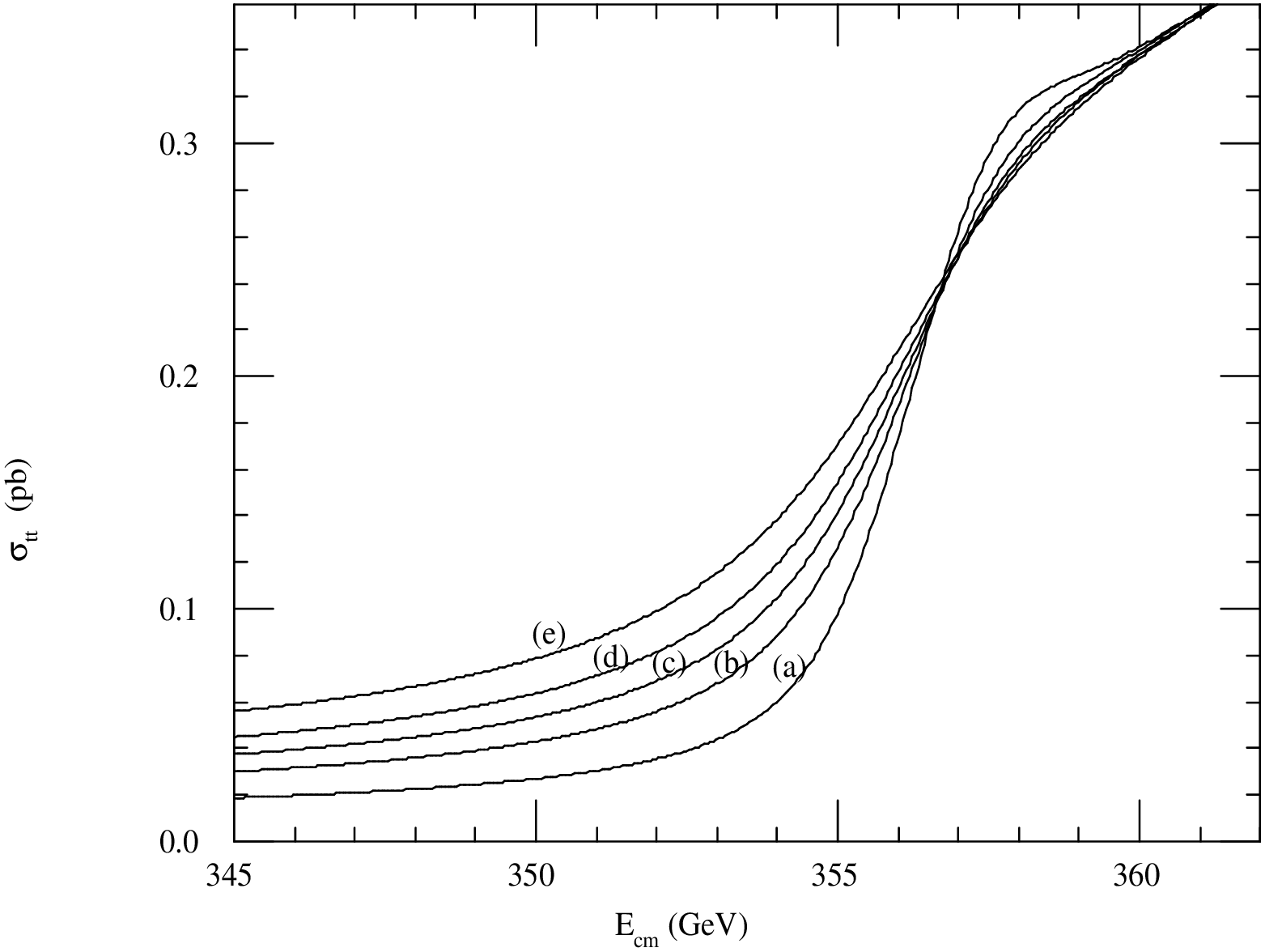}
{\par\noindent\eightrm
Fig.~8. Variation of threshold cross section with top width, 
as in Fig.~2, but
with initial-state radiation and all beam
effects included. A value $m_t=180$ GeV/c$^2$ is assumed.}

\vglue0.6cm
\leftline{\tenit 3.3. Momentum and Asymmetry Measurements }
\vglue0.4cm

As we have discussed, the lifetime of the toponium resonance
is determined by the first top quark to undergo weak decay, 
rather than by annihilation. This has the interesting implication
that the kinetic energy (or momentum) of the top daughter 
particles reflects the potential energy of the QCD interaction
before decay. Hence, a measurement of the momentum distribution
will be sensitive to $V_{\tenrm QCD}$ ({\it i.e.} Eq.~1) 
and $\alpha_s$. The 
theory$^{\RefPTH}$ and phenomenology$^{{\RefFUJII},{\RefPPH}}$ 
of this physics has been extensively studied. The observable
which has been used to characterize the distribution is the
momentum, $p_p$ at which the peak of the distribution occurs.
The value of $p_p$ at a given center of mass energy is indeed
found to be sensitive to $\alpha_s$.

It is intuitive that a measurement of the threshold cross section 
is the best method for measuring $m_t$. However, because of the
correlation between $m_t$ and $\alpha_s$ from the cross section,
as shown in Fig.~10b, it would be useful to have the momentum 
measurement be relatively insensitive to $m_t$. If one defines
the scan energy to be with respect to threshold, that is
$E= \sqrt{s} - 2m_t$, then this immediately introduces an
unwanted dependence on the knowledge of $m_t$. This is a
statement of experimental, as well as theoretical, uncertainty,
since the theoretical connection between the threshold shape and
$m_t$ is not necessarily exact within an offset of a few hundred 
MeV/ c$^2$. It is better to measure the energy point with respect
to a well-defined feature of the experimental threshold curve,
as emphasized in Ref. ~\RefFUJII, in which it is suggested to
measure with respect to the position of the 1S resonance bump,
$\Delta E = \sqrt{s} - \sqrt{s_{1S}}$.  One expects the 
average top momentum to vary roughly as the reciprocal of the
Bohr radius, or as $\alpha_s m_t$. On the other hand, as we saw
earlier, the energy of the 1S resonance is proportional to
$-\alpha_s^2 m_t$. So we expect $p_p$ as a function of $\Delta E$
to be approximately independent of $m_t$, but to retain 
sensitivity to $\alpha_s$. In fact, this is borne out in Figs.~9a
and 9c. As the top width increases, the top decays occur at 
shorter distances, therefore shifting the momenta to larger
values. This can be seen in Fig.~9b. Therefore the momentum
spectra can be used, in conjunction with the cross section points,
to separate the measurements of $m_t$, $\alpha_s$, and 
$|V_{tb}|$, or $\Gamma_t$. For $\Delta E \approx 2$ GeV or 
greater, the use of $p_p$ was found$^{{\RefFUJII},{\RefEURO}}$ 
to be very insensitive to ISR and beam effects for both JLC
and TESLA designs.  

\vskip 0.3in
\epsfxsize=2.8in \epsfbox{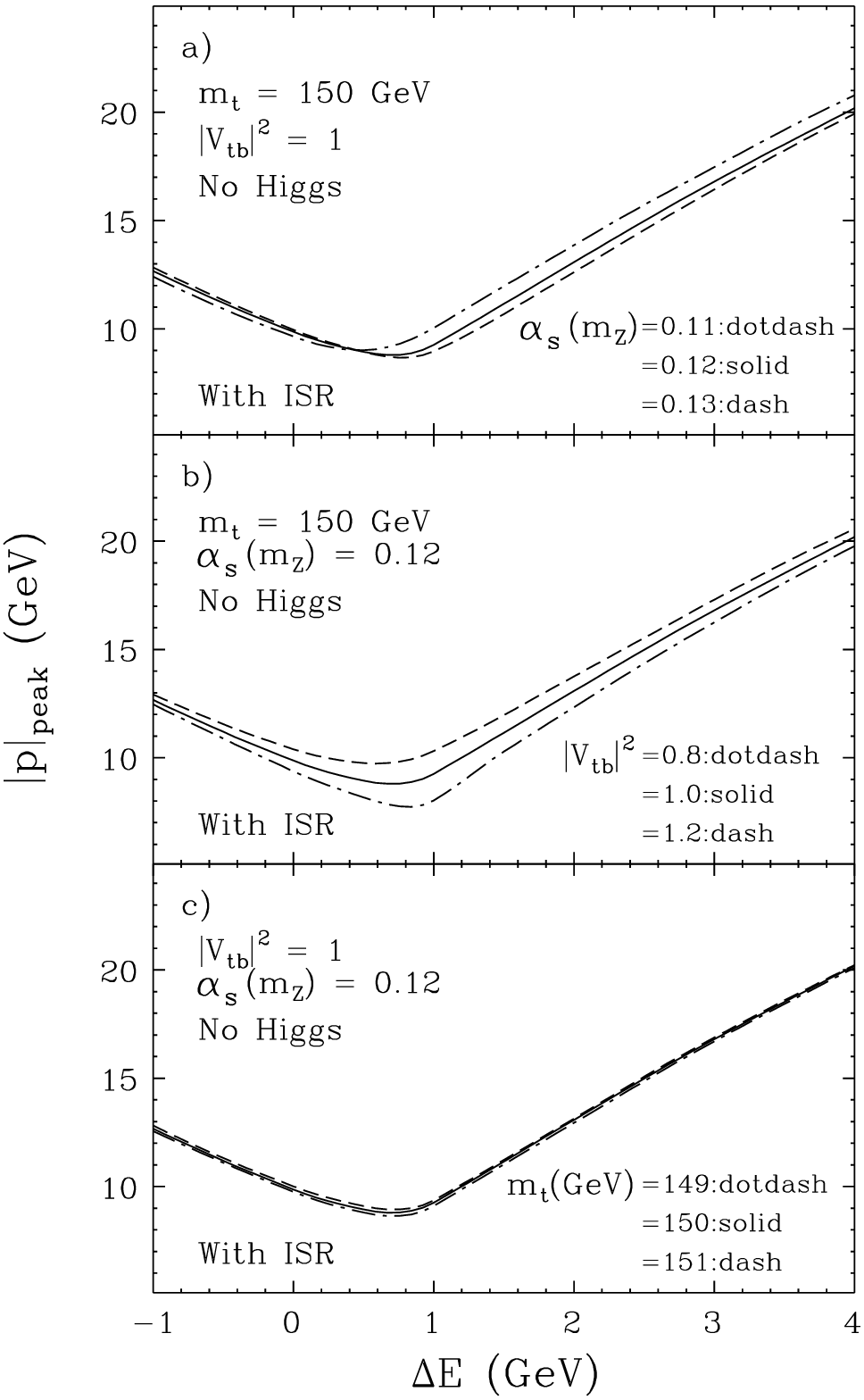}
\vglue13pt
{\par\noindent\leftskip 0.5in\rightskip 0.5in\eightrm
Fig.~9. Position of top peak momentum as a function of scan 
energy for different values of the following
physics input parameters (from Ref. ~{\RefFUJII}). 
(a) $\alpha_s$; (b) $ | V_{tb} | $; (c) $m_t$.  }

A quite different observable has been 
studied$^{{\RefSUMMUR},{\RefFUJII}}$ to help further 
pin down the physics parameters at threshold. Top is produced
symmetrically when produced in the 1S state.  
The vector coupling present with $Z$-$t$-$\bar{t}$ and 
$\gamma$-$t$-$\bar{t}$ can proceed to S and D-wave resonance
states. On the other hand, the axial-vector coupling present with
$Z$-$t$-$\bar{t}$ gives rise to P-wave resonance states. Hence, 
it is possible to produce interference between S and P-waves
which gives rise to a forward-backward asymmetry 
($A_{FB}$) proportional
to $\beta\cos\theta$. Because of the large width of the resonance
states, due to the large $\Gamma_t$, these states do overlap to
a significant extent, and a sizeable $A_{FB}$ develops. The
value of $A_{FB}$ varies from about 5\% to 12\% across the 
threshold, with the minimum value at about $\sqrt{s_{1S}}$. 
Since the top width controls the amount of S-P overlap, we expect
the forward-backward asymmetry to be a sensitive method for
measuring $\Gamma_t$.

\vglue0.6cm
\leftline{\tenit 3.4. Measurement of the Physics Parameters }
\vglue0.4cm

A number of studies have been carried out to simulate measurements
at $t-\bar{t}$ threshold. Figure 10a depicts a threshold scan$^{\RefFUJII}$
for which an integrated luminosity of 1 fb$^{-1}$ has been expended at
each of 10 values of nominal center-of-mass energy, $\sqrt{s}$. 
A value of $m_t=150$ GeV/c$^2$ was used. In this
case the 6-jet final state was selected, giving a branching fraction
times detection efficiency of 30\%. The physics background is measured
by the scan data taken below threshold. No beam polarization is assumed.
A fit of the data points to the theoretical cross section, including all
radiative and beam effects discussed above, results in a sensitivity for
the measurement of $m_t$ and $\alpha_s$ shown in Fig.~10b. The 
correlation between these two parameters, as discussed in Section 3.1,
is apparent. Even for the modest luminosity assumed here, the cross
section measurement gives quite good sensitivity to these quantities.
If no prior knowledge is assumed, errors for $m_t$ and $\alpha_s$ are
$200$ MeV/c$^2$ and $0.005$, respectively. Clearly, the value of
$\alpha_s(M_Z^2)$ is already known at LEP/SLC to this same
level of precision, implying that $m_t$ is considerably better determined,
approaching 100 MeV/c$^2$ in the limit where $\alpha_s(M_Z^2)$
is known exactly. This same cross section scan of 11 fb$^{-1}$ also
implies sensitivity to $\Gamma_t$ and the Yukawa coupling, $\lambda$,
of $0.2$ and $0.3$, respectively.
As described earlier, the measurement of the top
momentum and its forward-backward asymmetry can contribute
valuable additional information. The top momentum measusurement
alone produces a sensitivity corresponding to errors for
$\alpha_s$ and $|V_{tb}|$ of 0.002 and 0.04, respectively,
assuming an integrated luminosity of 100 fb$^{-1}$. The
the position of the 1S peak is assumed to be well known from the
cross section scan. The optimal energy for the momentum measurement
is $\Delta E = \sqrt{s} - \sqrt{s_{1S}}\approx 2$ GeV.
The $A_{FB}$ measurement at $\Delta E\approx 1$ GeV
provides an important crosscheck of the
total width and of $\alpha_s$, but requires more than twice the
luminosity relative to the momentum measurement to provide
similar sensitivity. We see that while the cross section scan
gives most of the sensitivity to the threshold parameters, the
momentum measurement in particular significantly increases sensitivity
to the top width measurement.
Similar threshold studies have been
presented$^{\RefLCWSF,\RefLCWSH}$ at previous meetings, with similar results.

\vskip 0.3in
\epsfxsize=2.8in \epsfbox{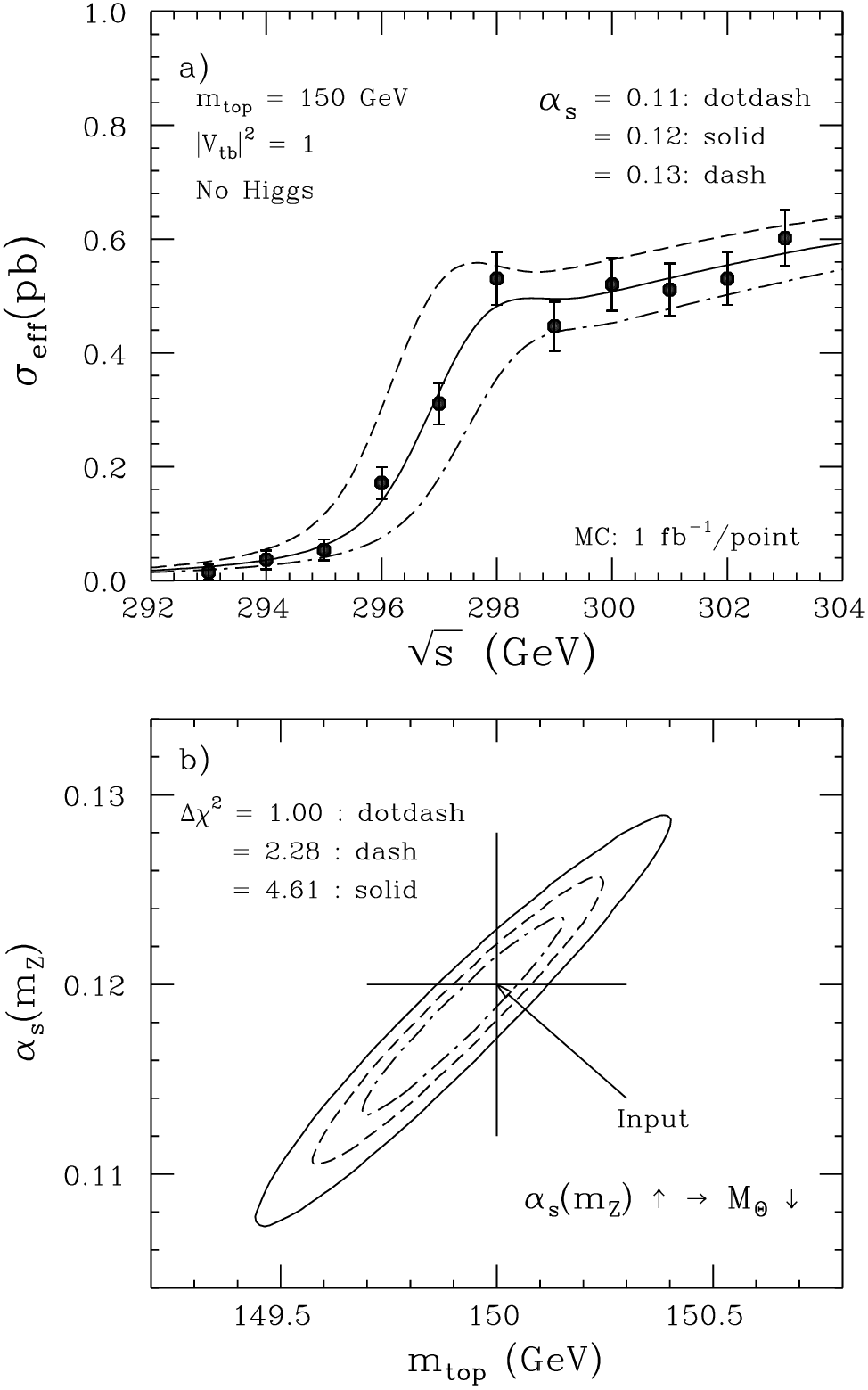}
\vglue13pt
{\par\noindent\leftskip 0.5in\rightskip 0.5in\eightrm
Fig.~10. (a) Top threshold scan from Ref.~{\RefFUJII}; (b) corresponding 
error ellipse
for $m_t$ and $\alpha_s$. A value for $m_t$ of 150 GeV/c$^2$ was assumed.}

At this meeting, new threshold studies by the European Working 
Group$^{\RefEURO}$ were presented assuming $m_t=180$ GeV/c$^2$
and TESLA beam parameters.
The cross section scan consisted of
10 points, each of 5 fb$^{-1}$ per point, with one of the points
below threshold to measure background. With an event selection consisting
of topological and mass cuts, the efficiency for 6-jet events was 33\%
with a signal to background ration of $5.5$. A 2-parameter fit to
$m_t$ and $\alpha_s$ yielded errors of 250 MeV/c$^2$ and $0.006$,
assuming no previous knowledge of these parameters. The single-parameter
sensitivities are 120 MeV/c$^2$ and $0.0025$ for $m_t$ and $\alpha_s$,
respectively. When the top momentum information for 4-jet$+\ell+\nu$
events from this same scan was included in the fit, the 2-parameter
fit errors improved to 200 MeV/c$^2$ and $0.005$ for 
$m_t$ and $\alpha_s$, respectively.
This analysis differs in principle from the one that
of Ref. ~{\RefFUJII} in that the scan energy used is $\sqrt{s}-2m_t$
rather than $\sqrt{s}-\sqrt{s_{1S}}$, as discussed in Section 3.3.
This has the efffect that the momentum measurement also produces
a strong correlation between $m_t$ and $\alpha_s$. However, since this
correlation is different from the one resulting from the cross section
scan, the two can still be disentangled. The 4-jet$+\ell+\nu$
events were also studied in an analysis of the forward-backward
asymmetry. These events were reconstructed with a $15\%$ efficiency
and a charge mis-identification rate of 3\%. With the same 50 fb$^{-1}$,
the $A_{FB}$ measurement, in conjunction with the cross section and
momentum measurements, gave a slight improvement in the $\Gamma_t$
sensitivity of $18\%$. 
However, it is clear, following 
Ref.~\RefFUJII and from Fig.~9, that threshold measurements optimized
for the $\Gamma_t$ measurement should be able to achieve a
sensitivity of 5--10\% for 50 fb$^{-1}$ and $m_t=180$ GeV/c$^2$.
The $m_t$-$\alpha_s$ error ellipse resulting
from these measurements is shown in Fig.~11.

\vskip 0.25in
\epsfxsize 3.0in \epsfbox[-100 0 400 525]{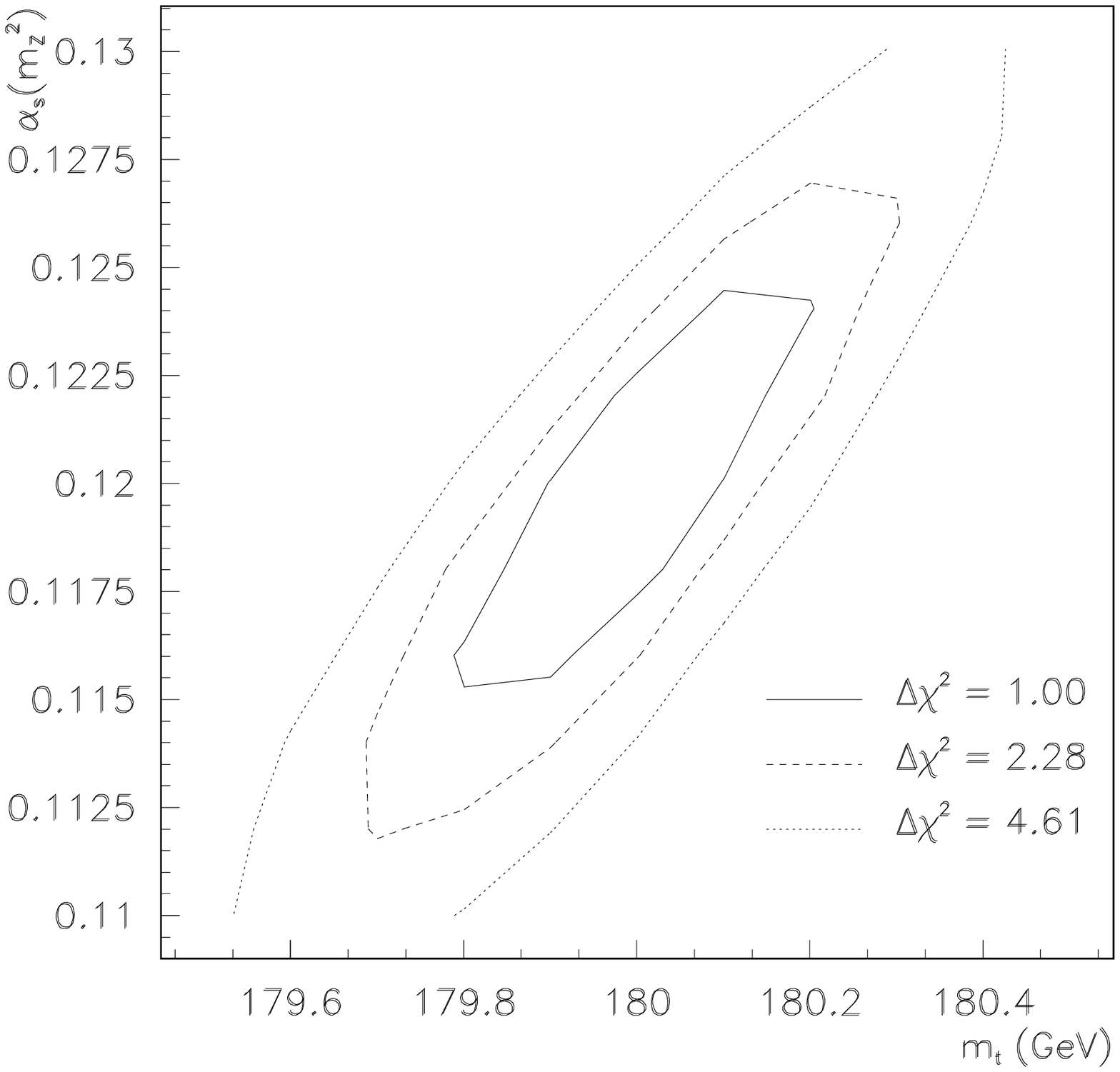}
{\par\noindent\leftskip 0.5in\rightskip 0.5in\eightrm
Fig.~11. Error ellipse for $m_t$ and $\alpha_s$ using cross section,
momentum peak, and forward-backward asymmetry information
from Ref.~\RefEURO. Mass and integrated
luminosity values of $m_t=180$ GeV/c$^2$ and 50 fb$^{-1}$, respectively, 
are assumed.}

\vfill\eject

\vglue0.6cm
\leftline{\tenit 3.5. Measurement of Luminosity Spectrum }
\vglue0.4cm

A unique experimental aspect of top threshold physics is the
necessity for knowledge of the integrated luminosity as a
function of the effective center of mass, or collision, energy.
A design year of data, after event selection, would result in an
event sample of $\sim 10^4$ events, and hence an overall statistical
error of roughly 1\%. Therefore, we would hope to measure the
luminosity expended at each scan point at a level approaching this
1\%. This requires knowledge of the luminosity spectrum, 
$d{\cal L}/d E_{cm}$, at a level commensurate with these errors. 
The nominal center-of-mass energy, $\sqrt{s}$, 
can in principle be measured using the same methodology as that
presently used at the SLC,$^{\RefWISRD}$ in which a precision
magnetic spectrometer is applied to the individual beams after
passing through the interaction point. An absolute energy 
measurement of $\sim 20$ MeV is achieved. At a FLC, this same
method would be applied, where one beam at a time would be turned
off in order to eliminate ISR and BS losses. One would want
to measure $\sqrt{s}$ with an error of $\sim 100$ MeV, which results
from simply scaling the SLC spectrometer from $M_Z$ to $2m_t$.

One might imagine that $\sqrt{s}$ including the energy loss due
to beamstrahlung could be measured with this same spectrometer
and with the beams put into collision. However, this is not
really the correct measurement, since for the physics collisions
the energy is luminosity sampled, whereas the average beamstrahlung
loss results from what is, in principle, different sampling. 
One would hope instead to measure a physics quantity which is
subject to exactly the same luminosity spectrum as that of
$t-\bar{t}$ production. This has been studied in Ref.~\RefMILLER,
where a program for carrying out a measurement of 
$d{\cal L}/d E_{cm}$ was proposed. The basic idea is to measure
a $2\rightarrow 2$ process, such as Bhabha scattering, where the
final-state acollinearity, $\theta_A$, can be measured in the
detector to good accuracy. The acollinearity is related
to the difference in the two beam energies, $\delta E$, according
to $\theta_A = (\delta E/E)\sin\theta$ for small $\theta_A$, where
$\theta$ is the scattering angle. The distribution in $\theta_A$
must then be related to $d{\cal L}/d E_{cm}$. The effects of ISR,
beamstrahlung, and single-beam energy spread on both $\theta_A$
and luminosity spectra can be calculated. This connection was
modelled in Ref.~\RefMILLER, where it was found that a measurement
error for $\theta_A$ at the level of $0.1\%$ is sufficient. 
Bhabha scattering has a high rate, of order $10^2$ times that for
top production, depending upon the angular region used, and hence
would be a good candidate for this measurement. 
Angular resolutions at the requisite level should be possible using
high-granularity electromagnetic calorimetry, for example 
using silicon strip readout layers.

It is important to confirm using real distributions and beam
parameters that this scheme can be carried out, and that the
requisite measurement errors are possible. The
beam parameters will change with time at some level, and hence the
contribution to the luminosity spectrum of beamstrahlung, 
which strongly depends on these parameters, will also change with time.
Of course, in the scheme outlined above, these variations will simply
be incorportated into the measured luminosity spectra by means of
the high-rate Bhabha scattering. However, the connection
between luminosity and acollinearity spectra using this technique
should be carefully checked using realistic calculations. 


\vglue0.6cm
\leftline{\tenbf 4. Top Couplings}
\vglue0.4cm

The motivation to examine non-standard top
couplings is clear, particularly in view of the large top mass.
There are a number of models of electroweak symmetry breaking for which
a heavy top quark is either an important or essential element.
In any case, as by far the heaviest known particle, it is important to examine
all top properties in as general and complete a manner as possible.
In this section, we investigate the sensitivity of a high-energy $e^+e^-$ linear
collider toward carrying out a general program of top quark coupling 
measurements. As we shall see, helicity amplitudes are
an important element of these studies, and the methods here make use of
the expected highly-polarized electron beam.

These studies make some basic assumptions of top properties. As discussed
before, with a mass of 180 GeV/c$^2$ it is
expected that top will decay before it hadronizes. Thus, in its decay,
$t\rightarrow bW$, the top spin information is directly transferred 
to the final state. This offers the unique opportunity 
to perform a conceptually clean helicity analysis by
means of top event reconstruction, in particular with the relatively 
clean final states available at a FLC, 
and in this way sensitively probe the top couplings. 
In the following sections, 
we briefly review the formalism for angular distributions
expected in top production and decay, and how these are affected by
non-standard couplings. We confine our discussion here to the study
of anomalous electroweak top couplings. A discussion of anomalous 
QCD coupling (chromomagnetic moments) has been presented$^{\RefPHIL}$
separately at this meeting.    

\vglue0.6cm
\leftline{\tenit 4.1. Angular Distributions and Couplings}
\vglue0.4cm  

The top neutral-current coupling can be generalized to the following
form for the Z-$t$-$\bar{t}$ or $\gamma$-$t$-$\bar{t}$ vertex factor:
$$\eqalignno{
{\cal M}^{\mu(\gamma ,Z)} &= e\gamma^\mu\left[ Q_V^{\gamma ,Z}F_{1V}^{\gamma ,Z}
+ Q_A^{\gamma ,Z}F_{1A}^{\gamma ,Z}\gamma^5  \right] \cr
&+ {{ie}\over{2m_t}}\sigma^{\mu\nu}k_\nu\left[
Q_V^{\gamma ,Z}F_{2V}^{\gamma ,Z}
+ Q_A^{\gamma ,Z}F_{2A}^{\gamma ,Z}\gamma^5 \right] ,&(6)\cr}
$$
which reduces to the familiar SM tree level expression when the form factors are
$ F_{1V}^\gamma = F_{1V}^Z = F_{1A}^Z = 1$, with all others zero. 
The quantities $Q_{A,V}^{\gamma ,Z}$ are the usual SM coupling constants:
$Q_V^{\gamma}=Q_A^{\gamma}={2\over 3}$, 
$Q_V^Z = (1-{8\over 3}\sin^2\theta_W)/(4\sin\theta_W\cos\theta_W)$,
and $Q_A^Z = -1/(4\sin\theta_W\cos\theta_W)$.
The non-standard couplings $F_{2V}^{\gamma ,Z}$ and $F_{2A}^{\gamma ,Z}$
correspond to electroweak magnetic and electric dipole moments, respectively. While
these couplings are zero at tree level in the SM,
the magnetic dipole coupling is expected to attain a value $\sim\alpha_s/\pi$
due to corrections
beyond leading order. On the other hand, the electric dipole analog term violates
CP and is expected to be zero in the SM through two loops.$^{\RefSUZUKI}$ 
Therefore, the search for a non-zero value is very interesting. Such
a non-standard coupling necessarily involves a top spin flip, 
hence the coupling is proportional to $m_t$.$^{\RefFC}$

\epsfxsize=4.3in \epsfbox{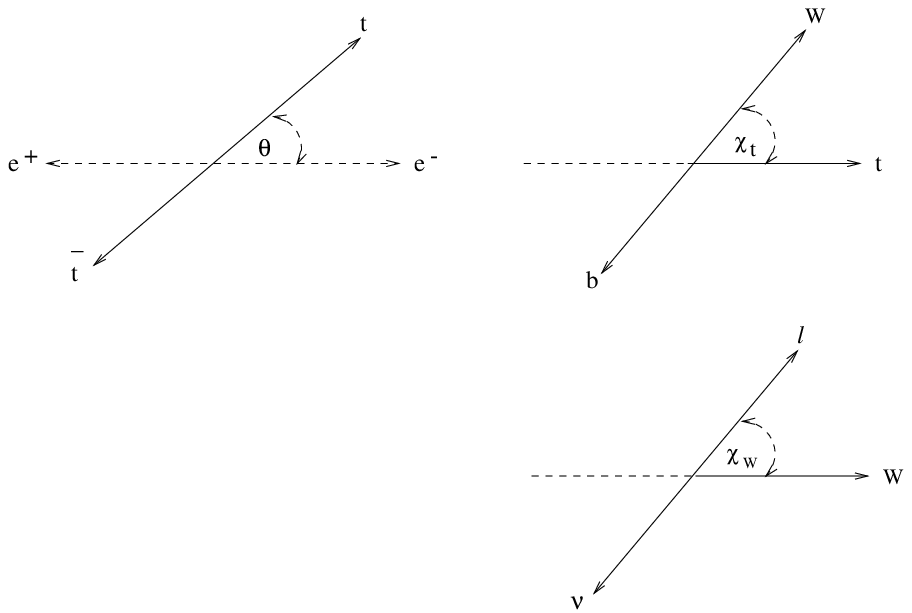}
{\par\noindent\eightrm
Fig.~12. Definitions of helicity angles. (a) Production angle $\theta$ in
$t\bar{t}$ proper frame; (b) $\chi_t$ measured in the top proper frame as shown;
and (c) $\chi_W$ in the W proper frame.}

\medskip
In terms of helicity amplitudes, the form factors obey distinct dependences
on the helicity state of $e^-$, $e^+$, $t$, and $\bar{t}$, which can be
accessed experimentally by beam polarization and the measurement of the
decay angles in the final state. These helicity angles can be defined as shown
in Fig.~12. The angle $\chi_W$ is defined in the 
W proper frame, so that the
W arrow represents its momentum vector in the limit of zero magnitude.
The analgous statement holds for the definition of $\chi_t$. 
Experimentally, all such angles, including the angles corresponding
to $\chi_t$ and $\chi_W$ for the $\bar{t}$ hemisphere, 
are accessible. This
requires full event reconstruction. Given the large number of constraints available
in these events, full reconstruction is entirely feasible,
and is discussed further in Section 4.2. However, to reconstruct $\theta$
one must also take into
account photon and gluon radiation in the event reconstruction. As discussed
earlier, photon radiation from the initial state is an important effect. But this
is to excellent approximation a purely longitudinal boost which can be handled
by demanding longitudinal momentum balance. Gluon radiation can be more subtle.
Jets remaining after reconstruction of $t$ and $\bar{t}$ can be due to gluon
radiation from $t$ or $b$, and the correct assignment must be decided based on
the kinematic constraints and the expectations of QCD. 
 
\bigskip
\par\noindent\leftskip 0.35in\rightskip 0.35in\baselineskip=10pt
{\eightrm Table 1. Dependence of the helicity amplitudes on
top production angle (see Fig.~12) and on the various neutral-current
form factors. The helicity components for $e^-$, $e^+$, $t$, and $\bar{t}$
define the helicity amplitude and are given in the first column, where
L$=-$ and R$=+$; the angular dependence is given by $f(\theta)$; and the
applicable form factors are given in the last column.}

\leftskip 0.in\rightskip 0.in\baselineskip=13pt
\smallskip
\begintable
$h(e^-),h(e^+),h(t),h(\bar{t})$ | $f(\theta)$ | form factors \crthick
$\pm$,$\mp$,$-$,$-$ | $\sin\theta$ | $F_{1V}$, $F_{2V}$, $F_{2A}$ \cr
$\pm$,$\mp$,$-$,$+$ | $1+\cos\theta$ | $F_{1V}$, $F_{1A}$ \cr
$\pm$,$\mp$,$+$,$-$ | $1-\cos\theta$ | $F_{1V}$, $F_{1A}$ \cr
$\pm$,$\mp$,$+$,$+$ | $\sin\theta$ | $F_{1V}$, $F_{2V}$, $F_{2A}$ 
\endtable

\medskip

The dependencies of the neutral current couplings to measurement of the
distribution of the
production angle $\theta$ are outlined in Table 1. The initial state helicity
is defined by the beam polarization, and the final state is determined by
measurement of the angular distributions. The explicit formulae for the form
factors can be found in Ref.~\RefYUAN.
In the case of the production angle
$\theta$, the SM expectations$^{\RefMPCS}$
are given in Fig.~13 for the various $t\bar{t}$
helicity combinations and for left and right-hand polarized electron beam.
We see, for example, that for left-hand polarized electron beam, top quarks
produced at forward angles are predominantly left handed, while 
forward-produced top quarks
are predominantly right handed when the electron beam is right-hand polarized.
These helicity amplitudes combine to produce the following general form
for the angular distribution:$^{\RefYUAN}$
$$ {d\sigma\over{d\cos\theta}} = {\beta_t\over{32\pi s}}\left[
c_0\sin^2\theta + c_+(1+\cos\theta)^2 + c_-(1-\cos\theta)^2 \right] ,\eqno(7) $$
where $c_0$ and $c_\pm$ are functions of the form factors.
The helicity structure of the event is highly constrained by
beam polarization and production angle. Clearly, the measurement of production
angle from event reconstruction, as well as the beam polarization, powerfully
constrain any non-standard contibutions to the form factors. 
 
\medskip

\epsfxsize=3.5in \epsfbox[90. 310. 400. 680.]{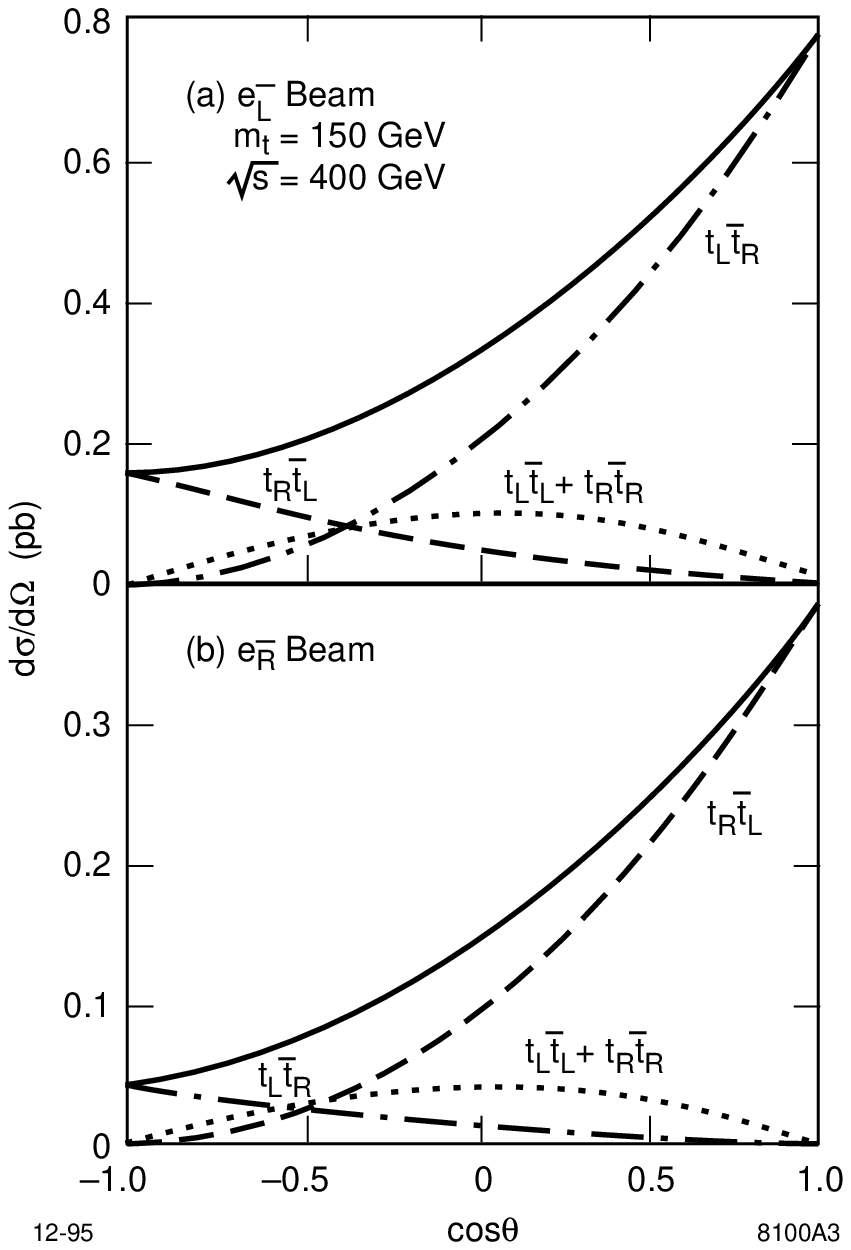}
{\par\noindent\eightrm
Fig.~13. Production angle for $t\bar{t}$ for the possible final-state helicity
combinations, as indicated, for (a) left-polarized electrons, and (b) right-polarized
electrons. The complete cross sections are the solid curves.}

\medskip
For the top charged-current coupling we can write the $W$-$t$-$b$ vertex factor as
$$
{\cal M}^{\mu,W}={g\over\sqrt{2}}\gamma^\mu\left[ P_L F_{1L}^W + P_R F_{1R}^W \right]
+ {{ig}\over{2\sqrt{2}\,m_t}}\sigma^{\mu\nu}k_\nu\left[
 P_L F_{2L}^W + P_R F_{2R}^W \right] ,\eqno(8)
$$
where the quantities $P_{L,R}$ are the left-right projectors.
In the SM we have $F_{1L}^W = 1$ and all others zero. The form factor $F_{1R}^W$
represents a right-handed, or $V+A$, charged current component.
The form of the angular distribution expected for each helicity combination 
at the $W$-$t$-$b$ vertex factor is given in Table 2, as well as 
the relevant form factor. Similarly, the angular dependence for the W decay 
is given in Table 3, where the SM couplings are assumed in this case.
Note that, as mentioned earlier, the case where the W is longitudinally
polarized is particularly relevant for heavy top.

\medskip
\par\noindent\leftskip 0.35in\rightskip 0.35in\baselineskip=10pt
{\eightrm Table 2. Dependence of the helicity amplitudes on the helicity angles,
as defined in Fig.~12, and on the charged-current form factors. The helicity
states for $t$ and $b$ are indicated by $\pm$, as in Table 1. The longitudinal
polarization state of W is indicated by $0$.}

\leftskip 0.in\rightskip 0.in\baselineskip=13pt
\smallskip
\begintable
$h(t),h(b),h(W)$ | $f(\chi_t)$ | form factors \crthick
$-$,$-$,$-$ | $\cos\chi_t/2$ | $F_{1L}$, $F_{2R}$ \cr
$-$,$-$,$0$ | $\sin\chi_t/2$ | $F_{1L}$, $F_{2R}$ \cr
$+$,$-$,$-$ | $\sin\chi_t/2$ | $F_{1L}$, $F_{2R}$ \cr 
$+$,$-$,$0$ | $\cos\chi_t/2$ | $F_{1L}$, $F_{2R}$ \cr
$-$,$+$,$+$ | $\sin\chi_t/2$ | $F_{1R}$, $F_{2L}$ \cr
$+$,$+$,$+$ | $\cos\chi_t/2$ | $F_{1R}$, $F_{2L}$ \cr
$-$,$+$,$0$ | $\cos\chi_t/2$ | $F_{1R}$, $F_{2L}$ \cr
$+$,$+$,$0$ | $\sin\chi_t/2$ | $F_{1R}$, $F_{2L}$
\endtable

\bigskip
\par\noindent\leftskip 0.35in\rightskip 0.35in\baselineskip=10pt
{\eightrm Table 3. Helicity angle dependence for W decay, as defined in Fig.~12.
The longitudinal polarization state of W is indicated by $0$.}

\leftskip 0.in\rightskip 0.in\baselineskip=13pt
\smallskip
\begintable
$h(W)$ | $f(\chi_W$ \crthick
$-$ | $\sin^2\chi_W/2$ \cr
$+$ | $\cos^2\chi_W/2$ \cr
$0$ | $\sin\chi_W$
\endtable

\vglue0.6cm
\leftline{\tenit 4.2 Form Factor Analyses}
\vglue0.4cm 

Three analyses of form factor measurement were presented in the parallel session.
These analyses are typical of those in the literature.
Cuypers$^{\RefFC}$ designed an analysis which is specifically sensitive to
the electroweak dipole moments, $F_{2A}^{\gamma ,Z}$. Two CP odd observables
are used: $(\vec{p}_b\times\vec{p}_{\bar{b}})\cdot\hat{z}$ and  
$(\vec{p}_b +\vec{p}_{\bar{b}})\cdot\hat{z}$, where $\hat{z}$ is along the 
incoming $e^+$.
The first observable is CPT even and probes the real part of the dipole moment,
whereas the second is CPT odd and probes the imaginary part.
The results show, for example, that for an integrated luminosity of 20 fb$^{-1}$ at
$\sqrt{s}=750$ GeV, then the 95\% CL limits on the real part of the dipole 
moment, $\Re(F_{2A}^Z)$, correspond to a dipole moment
of about $10^{-19}$ e-m. Efficiencies for $b$ and $W$ tagging of 10\% were
assumed. Electron beam polarization is an important element
of this analysis, and allows an increased sensitivity relative to similar
analyses performed without the assumption of beam polarization.

\centerline{
\epsfxsize=2.4in \epsfbox{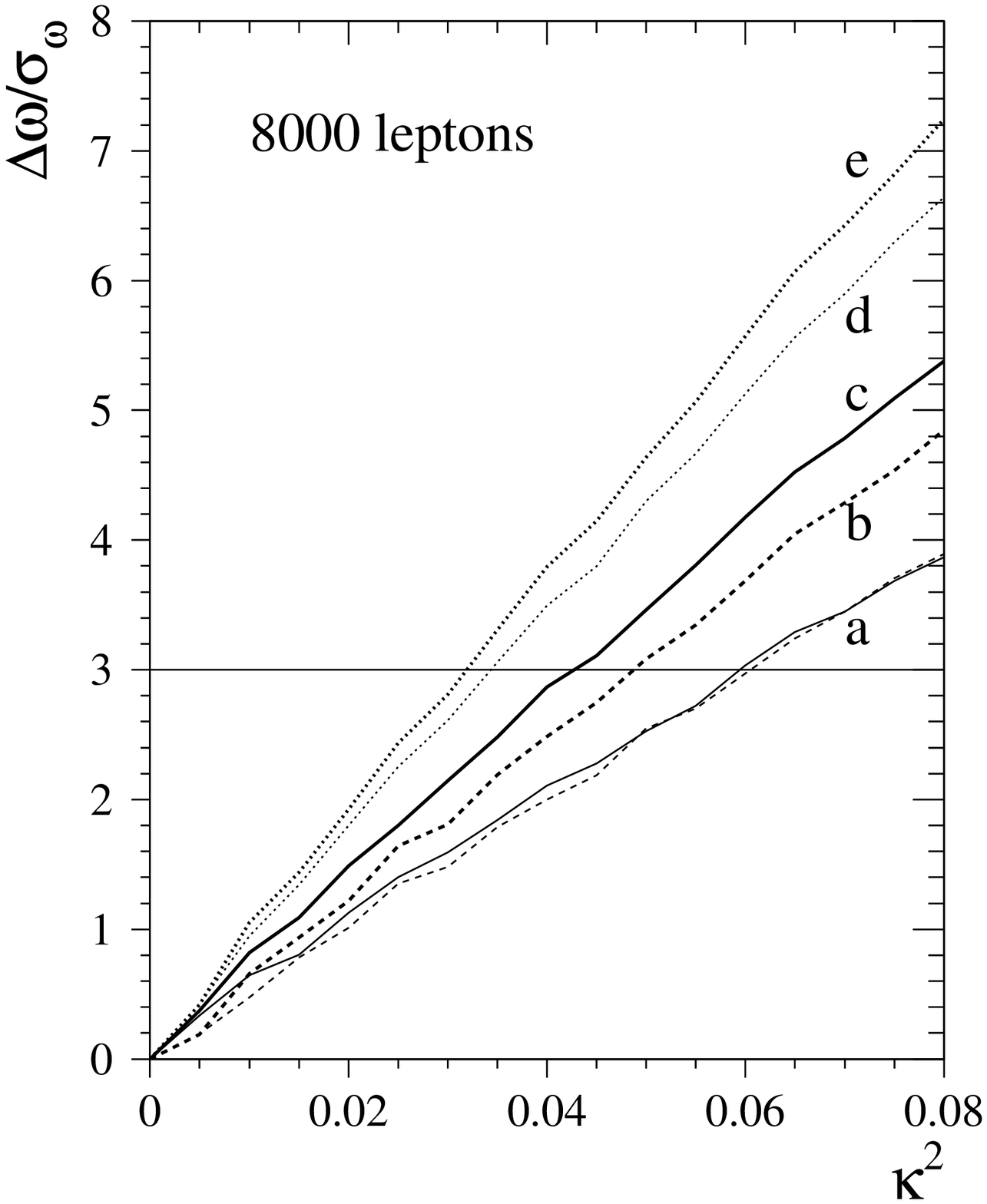}
}
{\par\noindent\eightrm
Fig.~14. Sensitivity to the measurement of an anomalous top charged-current
coupling from Ref.~\RefEURO. The curves $a$ through $e$ 
represent including incrementally more information in the event reconstruction.} 

\medskip
Martinez presented$^{\RefEURO}$ two form factor analyses, where the emphasis
was on examining the effects of including full Monte Carlo event generation
and realistic detector resolutions. In the first analysis, the neutral-current
couplings were examined by searching for a non-zero term proportional to 
$\sin^2\theta$ (see Eq.~(5) and Table 1). 
For $\sqrt{s}=500$, 50 fb$^{-1}$, and no
beam polarization, it was found that such a non-standard coupling could be
constrained to zero with a 68\% CL limit
of $1.5\%$ (relative to SM coupling) for
an ideal detector, increasing to $\approx 4\%$ after detection efficiency,
detector resolution, $t$--$\bar{t}$ mis-asignment,
and backgrounds were included.
An analysis of the charged-current coupling was performed at $t-\bar{t}$
threshold assuming 100 fb$^{-1}$ of data with no polarization. Threshold
was chosen so that the lab and center of momentum frame lepton momenta
from top decay are not very different. In this case, since the top spin is
oriented along the beam line, the helicity angle $\chi_t$ is effectively
measured directly in the laboratory frame once the $W\rightarrow \ell\nu$
is reconstructed. In Fig.~14 the sensitivity for measuring the quantity
$\kappa$ is displayed, where the SM couplings are modified according to
$ g_v = (1+\kappa)/\sqrt{1+\kappa^2}$ and $ g_a = (-1+\kappa)/\sqrt{1+\kappa^2}$.
The sensitivity is expressed in terms of the deviation from the SM by
a single quantity $\omega$ divided
by its measurement error, $\sigma_\omega$. As more information is included,
the sensitivity increases, as indicated by the curves $a$--$e$. It would
be interesting to compare the sensitivity using this technique at threshold
with one well above threshold which relies upon event reconstruction and
constraints.

\vglue0.6cm
\leftline{\tenit 4.2.1 Full-Event Analysis}
\vglue0.4cm 

We now discuss in more detail an analysis which can be applied in a
general way to the study of top couplings.
The results presented are limited in extent
and are of a preliminary nature, but the methodology readily
allows including more experimental detail. We use $m_t = 180$ GeV/c$^2$.
The only center-of-mass
energy considered here is 500 GeV. It should be repeated at higher energy.
We consider an integrated luminosity
of 10 fb$^{-1}$, which is quite reasonable given a typical design 
luminosity of $5\times 10^{33}$ cm$^{-2}$s$^{-1}$.
The decays are assumed to be $t\rightarrow bW$
followed by $WW\rightarrow \ell\nu\,qq^\prime$. Now,
since the top production and decay information is correlated, it
is possible to combine all relevant observables to ensure maximum sensitivity
to the couplings. In this study, a likelihood function is used to combine
the observables. The key tool for this study is the Monte Carlo generator
developed by Schmidt,$^{\RefSCHMIDT}$ which includes $t\bar{t}(g)$ production
to ${\cal O}(\alpha_s)$.  No hadronization is performed.  
Most significantly, the Monte Carlo correctly
includes the helicity information at all stages.

In general, one needs to distinguish $t$ from $\bar{t}$. The most
straightforward method for this is to demand that at least one of the W
decays be leptonic, and to use the charge of the lepton as the tag.
One might imagine using other techniques, for example with topological 
secondary vertex detection one could try to distinguish $b$ from $\bar{b}$.
Also, if the neutral-current couplings are determined and the charged-current
coupling is being studied in more detail, one could then use the polarization
and production angle to tag $t$ versus $\bar{t}$, as indicated by Fig.~13.
However, here it is simply assumed that one W decays hadronically and the
other W decays to electron or muon: 
$$t\bar{t}\rightarrow b\bar{b}WW \rightarrow 
b\bar{b}q\bar{q}^\prime\ell\nu ,\eqno(9)$$
where $\ell=e,\mu$. The branching fraction for this decay chain is 
$24/81$. The top decay products, including any jets due to 
hard gluon radiation, must be correctly assigned with good probability
in order to carry out this analysis. The correct assignments are rather
easily arbitrated using the W and top mass constraints. 
The effects of initial-state radiation and beamstrahlung give
rise to events in which the $t\bar{t}$ longitudinal momentum is unbalanced. 
This implies that additional information must be included in order to determine
the longitudinal momentum, $p_z^\nu$, of the neutrino
in $W\rightarrow\ell\nu$. In fact, Ladinsky and Yuan have shown$^{\RefYUAN}$ that
the $m_W$ and $m_t$ constraints determine $p_z^\nu$, with the correct sign, 
with an efficiency of about 70\%. They also show, indeed, that events
with doubly-leptonic W decays can also be correctly reconstructed with good
efficiency. For simplicity, we ignore this additional event fraction 
(BR$=4/81$ assuming $\ell=e,\mu$).

The Schmidt Monte Carlo is used to generate $t\bar{t}(g)$ with full 
${\cal O}(\alpha_s)$ corrections. Simple, phenomenological detection resolution
functions are then applied to the decay chain of Eq. (4). It is assumed that
the leptons and quarks
are measured with an energy resolutions of $\Delta E/E=0.15/\sqrt{E(GeV)}$
and $\Delta E/E=0.40/\sqrt{E(GeV)}$, respectively. 
The direction of quark momenta is
assumed to be precisely determined. The missing energy is then
assumed to have a resolution which is the quadrature sum of the lepton
and quark measurement errors. It is important to check the sensitivity of these
obvious over-simplifications to the results by applying experimental resolution
functions to fully hadronized events. Nonetheless, it is likely that the most
important effect of full event simulation may instead be due to jet 
mis-identification in these rather complicated 4(or 5)-jet final states.
In fact, we determine the RMS spread of the reconstructed top mass
distribution to be $\approx 8$ GeV, which is not very different from the
results based on simulations$^{\RefLCWSH}$ which include jet fragmentation.
It will also be interesting to apply what is expected to be highly efficient
b-jet tagging in order to reduce jet combinatorial inefficiency.
The acceptance is conservatively assumed to be zero for angles within $10^\circ$
of the beamline (forward and backward) due to the dead-cone masking. This
represents an inefficiency of only 3\%. Electron beam polarization is assumed
to be $\pm 80\%$.


Once the events are fully reconstructed, the resulting helicity 
angles (see Fig.~12) are then used to form a likelihood which is the square
of the theoretical amplitude for these angles given an assumed set of 
form factors. The likelihood is usually examined while a single form factor
value is varied from its nominal SM value. A typical result is given in Fig.~15,
in this case for a hypothetical right-hand $W$-$t$-$b$ coupling. The overall
efficiency of the analysis, including branching fractions, reconstruction
efficiency, and acceptance, is about 18\%.
This implies an event sample
of about 1400 events for 10 fb$^{-1}$ of luminosity with $\sqrt{s}=500$ GeV
and $m_t = 180$ GeV/c$^2$.

In the case shown in Fig.~15,
the use of beam polarization produces an obvious increase in sensitivity, as
evidenced by the steeper likelihood curve.
In many cases, because the helicity structure of these 
events is highly constrained by the Standard Model,
the polarization information is often seen to be formally redundant.
However, several points should be made. First,
an additional powerful experimental tool is generally found to be required,
rather than redundant, once all of the real-world uncertainties are included.
In addition, it may be the case that more than a single coupling is found
to exhibit an apparent deviation from its nominal SM value. Polarization will
likely allow the effects to be separated. A left-hand polarized electron beam
will always give a statistical advantage in the SM for $t\bar{t}$ production.
The cross sections are 680 fb, 360 fb, and 520 fb for 80\% left-polarized beam,
unpolarized beam, and 80\% right-polarized beam, respectively. In addition, as
mentioned previously, the fact that the $W^+W^-$ background goes away completely
in the limit of a fully right-polarized beam implies that its level is subject
to direct verification in the data.

\medskip
\epsfysize=3.3in \epsfbox[0 0 791 612]{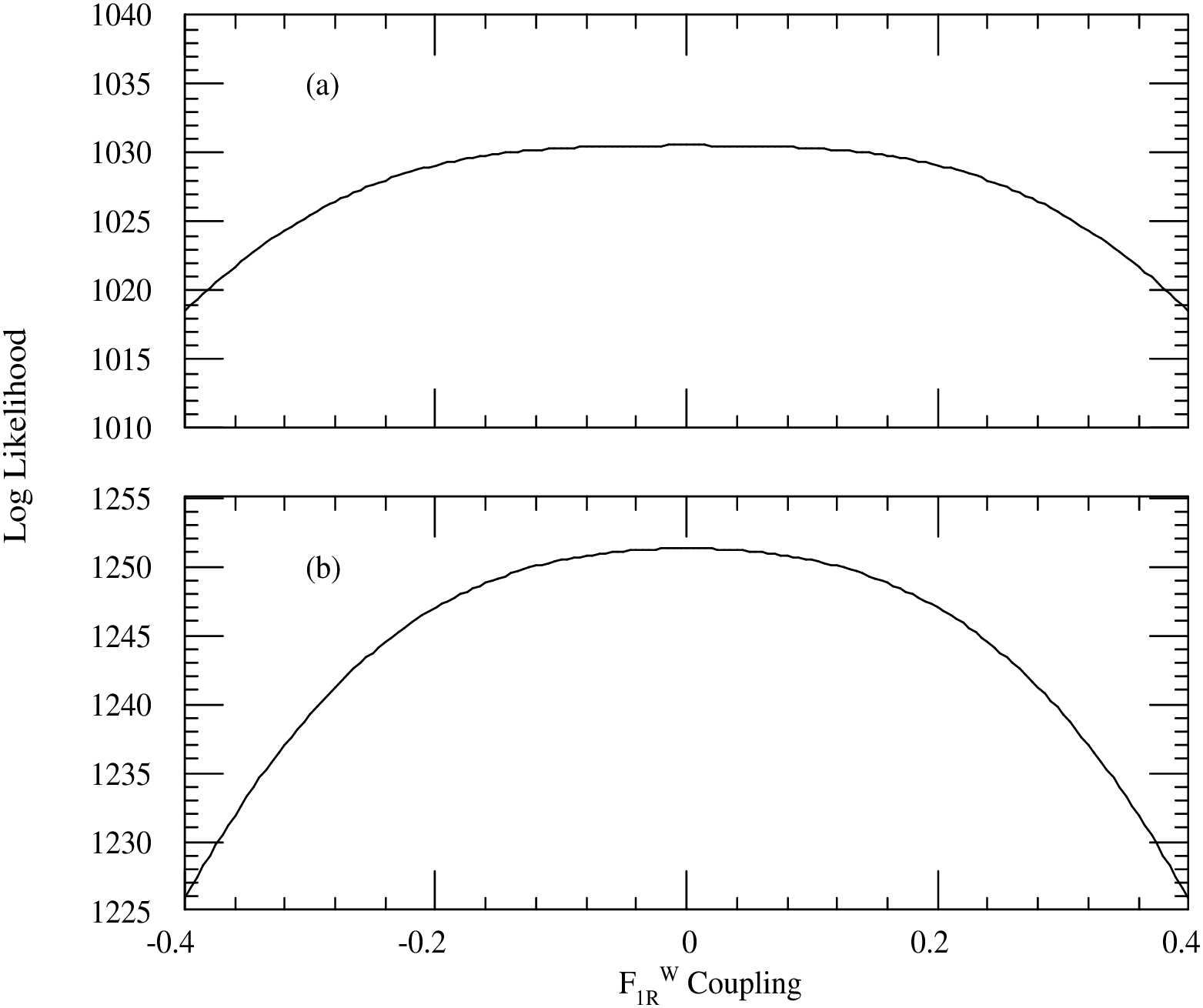}
{\par\noindent\eightrm
Fig.~15. An example of the variation of the logarithm of the likelihood funtion as a
function of coupling strength. In this case the coupling is a hypothetical
right-handed top charged-current coupling, $F_{1R}^W$, for 
(a) unpolarized beams, and (b) an $80\%$ left-polarized electron beam.}

\bigskip
Table 4 shows some of the results of this analysis. We see that even
with a modest integrated luminosity of 10 fb$^{-1}$ at $\sqrt{s}=500$ GeV,
the sensitivity to the form factors is quite good, at the level of
5--10\% relative to SM couplings. In terms of real units, 
the 90\% CL limits for $F_{2A}^Z$ of $\pm 0.15$, for example, 
correspond to a $t$-$Z$ electric dipole moment of 
$\sim 8\times 10^{-20}$ e-m. This type of analysis, which makes
use of the full event information, should be a very general and
powerful probe of the couplings. However, it is important to include 
the effects of background in such analyses. As shown in Ref.~\RefEURO,
this is unlikely to cause huge effects, and a larger data set can largely
compensate. In fact, the assumed sample of 10 fb$^{-1}$
is quite modest, especially given that no special beam energy is required,
and one would expect the quoted errors to scale approximately with statistical
error until the limits are very much smaller.

\medskip\par\noindent\leftskip 0.35in\rightskip 0.35in\baselineskip=10pt
{\eightrm Table 4. A sample of the preliminary results from the global analysis
described in the text. The upper and lower limits of the couplings in their
departures from the SM values are given at 68\% and 90\% CL for 10 fb$^{-1}$
and $m_t=180$ GeV/c$^2$. All couplings,
with real and imaginary parts, can be determined in this way. The right-handed
charged-current coupling is shown both for unpolarized and 80\% left-polarized
electron beam, whereas the other results assume 80\% left-polarized beam only.}

\medskip\leftskip 0.in\rightskip 0.in\baselineskip=13pt
\begintable
Form Factor | SM Value  | \multispan 2  Limits \nr
         | (Lowest Order)  | 68\% CL | 90\% CL\crthick
$F_{1R}^W (\cal P=0$) | 0 | $\pm 0.13$ | $\pm 0.18$ \cr
$F_{1R}^W (\cal P=80\%$) | 0 | $\pm 0.06$ | $\pm 0.10$ \cr
$F_{1A}^Z$ | 1 | 1$\pm 0.08$ | 1$\pm 0.13$ \cr
$F_{1V}^Z$ | 1 | 1$\pm 0.10$ | 1$\pm 0.16$ \cr
$F_{2A}^\gamma$ | 0 | $\pm 0.05$ | $\pm 0.08$ \cr
$F_{2V}^\gamma$ | 0 | $\pm 0.07$ | $^{+0.13}_{-0.11}$ \cr
$F_{2A}^Z$ | 0 | $\pm 0.09$ | $\pm 0.15$ \cr
$F_{2V}^Z$ | 0 | $\pm 0.07$ | $\pm 0.10$ \cr
$\Im (F_{2A}^Z)$ | 0 | $\pm 0.06$ | $\pm 0.09$ \endtable

\vfill\eject

\vglue0.6cm
\leftline{\tenbf 5. Direct Top Yukawa Coupling Measurement}
\vglue0.4cm

A fundamental tenet of the Standard Model is that the Higgs boson couples to
fermions with a strength proportional to the fermion mass, as given by
Eq.~4. This relationship clearly requires experimental scrutiny. As an example
of alternative relationships, in many SUSY models Eq.~4 is modified by a
factor $\cos\alpha/\sin\beta$, where $\alpha$ and $\beta$ have their usual
SUSY definitions. The large $m_t$ implies that the top Yukawa coupling may be
the only one which is accessible to experimental test. The sensitivity of
the top threshold shape to Higgs exchange was discussed in Section 3.1 as
a means toward Yukawa coupling measurement. Here we 
briefly discuss the possibilities
for directly measuring this coupling in open top production. 
Figure 16 indicates representative diagrams which contribute to the 
relevant processes. One assumes that the Higgs boson(s) would have already
been discovered, for example at LEP II, LHC, or FLC.

\epsfxsize=4.75in \epsfbox{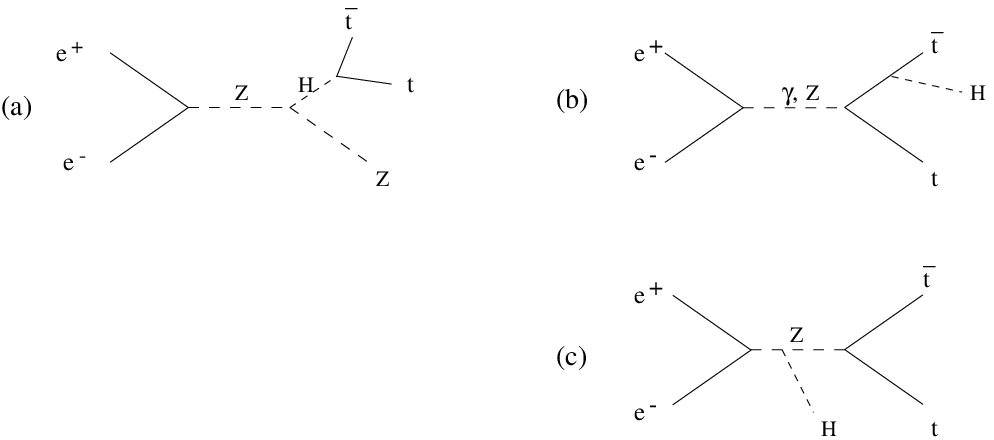}
{\par\noindent\leftskip 0.5in\rightskip 0.5in\eightrm
Fig.~16. Examples of 
processes discussed in the text for the study of 
the top-Higgs coupling. }

Figure 16a represents production of $Z^*H$, where the Higgs is sufficiently
massive to allow decay to $t\bar{t}$. This has been studied in Ref.~\RefTAUCHI,
where it is found that for $m_t=130$ GeV/c$^2$ and $M_H=300$ GeV/c$^2$
it is possible to reconstruct $\sim 30$ events with $M(t\bar{t})\approx M_H$
over a small background at the Higgs mass, and to measure
the Yukawa coupling to 10\% for 60 fb$^{-1}$ and $\sqrt{s}=600$ GeV. However,
with realistic values of $m_t$, the cross section for this process drops
rapidly, and would require extended running at high energy.

Rather than the heavy-Higgs, light-top scenario which is best for the
process of Fig.~16a, we now find the more interesting possibility that
of the so-called ``Higgs-strahlung'' represented by Fig.~16b, which is
important for a light-Higgs, heavy-top scenario. A study of this process
was presented at this meeting$^{\RefEURO}$ for $m_t=180$ GeV/c$^2$,
and is summarized here.
Sensitivity to this process for a given $M_H$ increases with $\sqrt{s}$.
A cross section of about 1 fb is attained for $M_H=100$ GeV/c$^2$ at 
$\sqrt{s}=500$ GeV and for $M_H=200$ GeV/c$^2$ at 1 TeV.
A perfect reconstruction would produce a 10\% or better measurement of the Yukawa
coupling for $M_H<240$ GeV/c$^2$, assuming an integrated luminosity of
50 fb$^{-1}$. While this seems promising, one must realize that these
events, consisting typically of 6 or 8 jets, are not trivially reconstructed.
And they must be separated from the background due to $t\bar{t}Z$ and $t\bar{t}$.
A complete simulation was attempted for the $t\bar{t}H\rightarrow 8$ jets signal.
It was found that due to its relatively high rate, the background from
$t\bar{t}$ was most troublesome, and a signal to background ratio $\sim 1$
was determined from the simulations. 

While this study is preliminary, it underscores the experimental challenge
in studying this important process. Since we expect Higgs to decay primarily
to $b\bar{b}$, then $t\bar{t}H$ events will consist of the final states
4 b-jet + 2-$\ell$ + 2-$\nu$ ($\sim 11\%$); 
4 b-jet + $qq^\prime$ + $\ell\nu$ ($44\%$); and
4 b-jet + 2-$qq^\prime$ ($45\%$). Clearly, efficient b-tagging is an important
tool for studying these events. And the subtleties associated with forming
correct multi-jet masses in this busy environment will be important to
study. 

It turns out that the Higgs-strahlung process is also sensitive to 
deviations from the Standard Model involving extended Higgs sectors. The
process represented by Fig.~16c also gives rise to the $t\bar{t}H$ final
state. Interference between this and the processes represented by Fig.~16b
is sensitive to extended Higgs sectors. In fact, Ref.~\RefCP shows that
in many two-Higgs doublet models, including those favored by SUSY, this
interference gives rise to large CP violation. They introduce CP-odd
observables, which given roughly 100 fb$^{-1}$ of data at $\sqrt{s}=800$
GeV, could produce an observable CP asymmetry in $t\bar{t}H$ events,
although the result depends on which parameters are chosen in the
two-Higgs doublet model.

\vfill\eject

\vglue0.6cm
\leftline{\tenbf 6. Summary}
\vglue0.4cm

The discovery of top, especially the large measured mass, has brought
the most important new input to the study of top physics at a
high-energy linear $e^+e^-$ collider. The main issues 
representing this physics has been covered, although to largely varying
levels of detail and subtlety, in previous conference proceedings and
journal articles. However, the large $m_t$ has produced a qualitative 
change in the outlook for top physics. For example, the top threshold
shape becomes less distinctive, with the corresponding measurement of
parameters less precise. But the overwhelming change brought upon by the
large mass is the sense that the top quark is truly special, and in fact
may well play a major role in electroweak symmetry breaking, either directly
or indirectly, or other physics at large mass. Hence, the measurement of
the large top Yukawa coupling and the various electroweak couplings have
taken on a new importance. 

At this meeting, there were presentations which updated threshold studies
for the measured $m_t$. These measurements will offer beautiful and
unique tests of QCD, as well as unsurpassed measurements of top mass and
width. A design year at a FLC at threshold would provide sensitivity to
$m_t$ and $\alpha_s$ at the level of 120 MeV/c$^2$ and 0.0025, respectively.
Similarly, the sensitivity to the total top decay width is roughly 10\%.
Accelerator and detector designs have become sufficiently stable to 
make possible calculations which incorporate the systematics associated
with luminosity spectra and backgrounds. This would allow
determination of the limiting systematic errors at threshold. For example,
is the measurement of beam energy, at the level of 100 MeV, really the
limiting systematic for $m_t$ measurement? With the large top mass, the
contribution of Higgs exchange at threshold should become measureable,
assuming that the systematics are under control.

Several presentations were given for direct Yukawa and electroweak coupling
measurements. The Yukawa coupling measurement via the Higgs-strahlung process 
looks
promising in principle, but in practice may present a significant challenge
to the experimentalist, which will be very interesting to pursue. A
complete set of measurements of 
electroweak gauge and dipole couplings of top to the
charged and neutral currents is possible at a FLC. A modest sample of
10 fb$^{-1}$ at $\sqrt{s}=500$ GeV typically constrains these couplings,
both real and imaginary parts, to within about 5--10\% of their lowest-order
Standard Model values. 

Besides including increasing levels of reality to these studies 
based upon better simulations of signals and backgrounds, a few experimental
techniques should prove generally applicable to these studies,
but until now have not been fully utilized. The existence
of electron beams with 80\% polarization already exist at SLC, and this
level of polarization, or better, should be easily achieved at a FLC. For
some measurements, for example of top couplings, the improvements with
polarized beam are integral. But for other analyses, at the very least, polarization
gives an increased top cross section (left-handed electrons) and dramatic
reduction of $W^+W^-$ background (right-handed electrons). The other 
demonstrated technique is precision vertex detection, where the small beam
size and small, stable interaction point within FLC detectors represents a
powerful asset which hopefully will be exploited in top physics
studies at future meetings.

\vglue0.6cm
\leftline{\tenbf 7. Acknowledgements}
\vglue0.4cm

I sincerely thank the local organizers of LCWS95 for a truly splendid job of
organization and for providing an enjoyable and stimulating environment. I
also thank the editors for their patience. I have benifited greatly from
helpful discussions with many colleagues, including K. Fujii, M. Peskin,
J. Jaros, M. Fero, C. Schmidt, D. Burke, and T. Barklow.

\vglue0.6cm
\leftline{\tenbf 8. References}
\vglue0.4cm

\medskip

\itemitem{\RefCDF.} F. Abe,{\it et al.} (CDF Collaboration), {\it Phys.
Rev. Lett.~} {\bf 74} (1995) 2626; \break
S. Abachi,{\it et al.} (D0 Collaboration), {\it Phys.
Rev. Lett.~} {\bf 74} (1995) 2632.

\itemitem{\RefLCWSF.} K. Fujii, in {\it Conference on Physics and Experiments
with Linear Colliders}, Saariselka, Finland, 1991.

\itemitem{\RefLCWSH.} P. Igo-Kemenes, in {\it Conference on Physics and 
Experiments with Linear Colliders}, Waikoloa, Hawaii, USA, 1993.

\itemitem{\RefPOL.} See K. Abe, {\it et al., Phys. Rev. Lett.~}{\bf 73}
(1994) 25, and references therein.

\itemitem{\RefTIMB.} T. Barklow, contribution to these proceedings.

\itemitem{\RefONETOP.} H.J. Schreiber, contribution to these proceedings.

\itemitem{\RefJKT.} M. Jezabek, J.H. Kuhn, and T. Teubner, {\it
Z. Phys.~}{\bf C56} (1992) 653.

\itemitem{\RefORR.} G. Jikia, {\it Phys. Lett.~}{\bf 257B} (1991) 196; \hfil\break
V.A. Khoze, L.H. Orr, and W.J. Stirling, {\it Nucl. Phys.}{\bf B378} (1992) 413.

\itemitem{\RefFUJII.} K. Fujii, T. Matsui, and Y. Sumino, {\it Phys. Rev.~} 
{\bf D50} (1994) 4341.

\itemitem{\RefCJSD.} C.J.S. Damerell, contribution to these proceedings.

\itemitem{\RefKF.} E.A. Kuraev and V.S. Fadin, {\it Sov. J. Nucl. Phys.~}
{\bf 41} (1985) 466.

\itemitem{\RefCHEN.} Pisin Chen, {\it Phys. Rev.~}{\bf D46} (1992) 1186.

\itemitem{\RefTHTH.} V. Fadin and V. Khoze, {\it JETP Lett.~}{\bf 46} (1987) 525
and {\it Sov. J. Nucl. Phys.~}{\bf 48} (1988) 309;\hfil\break
M. Strassler and M. Peskin, {\it Phys. Rev.~}{\bf D43} (1991) 1500;\hfil\break
M. Jezabek, J. Kuhn, and T. Teubner, {\it Z. Phys.~}{\bf C56} (1992) 653;
\hfil\break
M. Jezabek and T. Teubner, {\it Z. Phys.~}{\bf C59}, (1993) 669;\hfil\break
Y. Sumino, K. Fujii, K.Hagiwara, H. Murayama, and C.-K. Ng,
{\it Phys. Rev.}\hfil\break{\bf D47} (1993) 56.

\itemitem{\RefPESKIN.} M. Strassler and M. Peskin, 
{\it Phys. Rev.~}{\bf D43} (1991) 1500.

\itemitem{\RefFADIN.} V. Fadin and V. Khoze, {\it JETP Lett.}{\bf 46} 
(1987) 525 and {\it Sov. J. Nucl. Phys.~}{\bf 48} (1988) 309.

\itemitem{\RefHJK.} R. Harlander, M. Jezabek, and J.H. Kuhn, 
TTP-95-25, hep-ph/9506292, 1995.

\itemitem{\RefFRANKZ.} F. Zimmermann and T.O. Raubenheimer,
``Compensation of Longitudinal Nonlinearities in the
NLC Bunch Compressor'', SLAC-PUB-95-7020,
presented at {\it Micro Bunch}, Upton, N.Y., Sept. 1995.

\itemitem{\RefEURO.} M. Martinez,{it et al.,} 
(European Top Physics Group), contribution to these proceedings.

\itemitem{\RefPTH.} Y. Sumino, K. Fujii, K.Hagiwara, H. Murayama, 
and C.-K. Ng,{\it Phys. Rev.}\hfil\break{\bf D47} (1993) 56;\hfil\break
M. Jezabek, J. Kuhn, and T. Teubner, {\it Z. Phys.~}{\bf C56} (1992) 653.

\itemitem{\RefPPH.} P. Igo-Kemenes, M. Martinez, R. Miquel, and S. Orteu,
in {\it Conference on Physics and 
Experiments with Linear Colliders}, Waikoloa, Hawaii, USA, 1993.

\itemitem{\RefSUMMUR.}
H. Murayama and Y. Sumino, {it Phys. Rev.~}{\bf D47} (1993) 82.

\itemitem{\RefWISRD.} J. Kent, {\it et al.,} SLAC-PUB-4922, LBL-26977, 1989.

\itemitem{\RefMILLER.} N.M. Frary and D. Miller, DESY 92-123A, Vol. I, 1992,
p. 379.

\itemitem{\RefPHIL.} See P.N. Burrows, contribution to these proceedings.

\itemitem{\RefSUZUKI.} W. Bernreuther and M. Suzuki, {\it Rev. Mod. Phys.~} 
{\bf 63} (1991) 313.

\itemitem{\RefFC.} F. Cuypers, contribution to these proceedings.

\itemitem{\RefMPCS.} M.E. Peskin and C.R. Schmidt, in 
{\it Conference on Physics and Experiments
with Linear Colliders}, Saariselka, Finland, 1991.

\itemitem{\RefYUAN.} G.A. Ladinsky and C.-P. Yuan, {\it Phys.
Rev.~} {\bf D49} (1994) 4415; see also references therein.

\itemitem{\RefSCHMIDT.} C.R. Schmidt, SCIPP-95/14 (1995), hep-ph/9504434.

\itemitem{\RefTAUCHI.} K. Fujii, in contribution by P. Igo-Kemenes, 
in {\it Conference on Physics and 
Experiments with Linear Colliders}, Waikoloa, Hawaii, USA, 1993.

\itemitem{\RefCP.}
S. Bar-Shalom, {\it et al., Phys. Rev.~}{\bf D53} (1996) 1162.

\bye

%% file: tables.tex
%
\newbox\hdbox%
\newcount\hdrows%
\newcount\multispancount%
\newcount\ncase%
\newcount\ncols
\newcount\nrows%
\newcount\nspan%
\newcount\ntemp%
\newdimen\hdsize%
\newdimen\newhdsize%
\newdimen\parasize%
\newdimen\spreadwidth%
\newdimen\thicksize%
\newdimen\thinsize%
\newdimen\tablewidth%
\newif\ifcentertables%
\newif\ifendsize%
\newif\iffirstrow%
\newif\iftableinfo%
\newtoks\dbt%
\newtoks\hdtks%
\newtoks\savetks%
\newtoks\tableLETtokens%
\newtoks\tabletokens%
\newtoks\widthspec%
%
%
%
%
\tableinfotrue%
\catcode`\@=11
%
%
\def\tstrut{\vrule height3.1ex depth1.2ex width0pt}%
\def\and{\char`\&}
\def\tablerule{\noalign{\hrule height\thinsize depth0pt}}%
\thicksize=1.5pt
\thinsize=0.6pt
\def\thickrule{\noalign{\hrule height\thicksize depth0pt}}%
\def\ctr#1{\hfil\ #1\hfil}%
%
%
%
%
\tablewidth=-\maxdimen%
\spreadwidth=-\maxdimen%
\def\tabskipglue{0pt plus 1fil minus 1fil}%
%
%
\centertablestrue%
%
%
%
%
\parasize=4in%
\gdef\ARGS{########}
\gdef\headerARGS{####}
\def\@mpersand{&}
{\catcode`\|=13
\gdef\letbarzero{\let|0}
\gdef\letbartab{\def|{&&}}%
\gdef\letvbbar{\let\vb|}%
}
{\catcode`\&=4
\def\ampskip{&\omit\hfil&}
\catcode`\&=13
\let&0
\xdef\letampskip{\def&{\ampskip}}%
\gdef\letnovbamp{\let\novb&\let\tab&}
}
\def\begintable{
   \begingroup%
   \catcode`\|=13\letbartab\letvbbar%
   \catcode`\&=13\letampskip\letnovbamp%
   \def\multispan##1{
      \omit \mscount##1%
      \multiply\mscount\tw@\advance\mscount\m@ne%
      \loop\ifnum\mscount>\@ne \sp@n\repeat%
   }
   \def\|{%
      &\omit\widevline&%
   }%
   \ruledtable
}
\long\def\ruledtable#1\endtable{%
%
%
%
   \offinterlineskip
   \tabskip 0pt
   \def\widevline{\vrule width\thicksize}
   \def\endrow{\@mpersand\omit\hfil\crnorm\@mpersand}%
   \def\crthick{\@mpersand\crnorm\thickrule\@mpersand}%
   \def\crthickneg##1{\@mpersand\crnorm\thickrule
          \noalign{{\skip0=##1\vskip-\skip0}}\@mpersand}%
   \def\crnorule{\@mpersand\crnorm\@mpersand}%
   \def\crnoruleneg##1{\@mpersand\crnorm
          \noalign{{\skip0=##1\vskip-\skip0}}\@mpersand}%
   \let\nr=\crnorule
   \def\endtable{\@mpersand\crnorm\thickrule}%
   \let\crnorm=\cr
%
%
   \edef\cr{\@mpersand\crnorm\tablerule\@mpersand}%
   \def\crneg##1{\@mpersand\crnorm\tablerule
          \noalign{{\skip0=##1\vskip-\skip0}}\@mpersand}%
   \let\ctneg=\crthickneg
   \let\nrneg=\crnoruleneg
   \the\tableLETtokens
%
%
   \tabletokens={&#1}
%
%
   \countROWS\tabletokens\into\nrows%
   \countCOLS\tabletokens\into\ncols%
%
%
   \advance\ncols by -1%
   \divide\ncols by 2%
   \advance\nrows by 1%
%
%
   \iftableinfo %
      \immediate\write16{[Nrows=\the\nrows, Ncols=\the\ncols]}%
   \fi%
%
%
   \ifcentertables
      \ifhmode \par\fi
      \line{
      \hss
   \else %
      \hbox{%
   \fi
      \vbox{%
         \makePREAMBLE{\the\ncols}
         \edef\next{\preamble}
         \let\preamble=\next
         \makeTABLE{\preamble}{\tabletokens}
      }
      \ifcentertables \hss}\else }\fi
   \endgroup
   \tablewidth=-\maxdimen
   \spreadwidth=-\maxdimen
}
\def\makeTABLE#1#2{
   {
   \let\ifmath0
   \let\header0
   \let\multispan0
%
%
   \ncase=0%
   \ifdim\tablewidth>-\maxdimen \ncase=1\fi%
   \ifdim\spreadwidth>-\maxdimen \ncase=2\fi%
   \relax
%
   \ifcase\ncase %
      \widthspec={}%
   \or %
      \widthspec=\expandafter{\expandafter t\expandafter o%
                 \the\tablewidth}%
   \else %
      \widthspec=\expandafter{\expandafter s\expandafter p\expandafter r%
                 \expandafter e\expandafter a\expandafter d%
                 \the\spreadwidth}%
   \fi %
   \xdef\next{
      \halign\the\widthspec{%
      #1
      \noalign{\hrule height\thicksize depth0pt}
      \the#2\endtable
%
      }
   }
   }
   \next
}
\def\makePREAMBLE#1{
   \ncols=#1
   \begingroup
   \let\ARGS=0
   \edef\xtp{\widevline\ARGS\tabskip\tabskipglue%
   &\ctr{\ARGS}\tstrut}
   \advance\ncols by -1
   \loop
      \ifnum\ncols>0 %
      \advance\ncols by -1%
      \edef\xtp{\xtp&\vrule width\thinsize\ARGS&\ctr{\ARGS}}%
   \repeat
   \xdef\preamble{\xtp&\widevline\ARGS\tabskip0pt%
   \crnorm}
   \endgroup
}
\def\countROWS#1\into#2{
   \let\countREGISTER=#2%
   \countREGISTER=0%
   \expandafter\ROWcount\the#1\endcount%
}%
\def\ROWcount{%
   \afterassignment\subROWcount\let\next= %
}%
\def\subROWcount{%
   \ifx\next\endcount %
      \let\next=\relax%
   \else%
      \ncase=0%
      \ifx\next\cr %
         \global\advance\countREGISTER by 1%
         \ncase=0%
      \fi%
      \ifx\next\endrow %
         \global\advance\countREGISTER by 1%
         \ncase=0%
      \fi%
      \ifx\next\crthick %
         \global\advance\countREGISTER by 1%
         \ncase=0%
      \fi%
      \ifx\next\crnorule %
         \global\advance\countREGISTER by 1%
         \ncase=0%
      \fi%
      \ifx\next\crthickneg %
         \global\advance\countREGISTER by 1%
         \ncase=0%
      \fi%
      \ifx\next\crnoruleneg %
         \global\advance\countREGISTER by 1%
         \ncase=0%
      \fi%
      \ifx\next\crneg %
         \global\advance\countREGISTER by 1%
         \ncase=0%
      \fi%
      \ifx\next\header %
         \ncase=1%
      \fi%
      \relax%
      \ifcase\ncase %
         \let\next\ROWcount%
      \or %
         \let\next\argROWskip%
      \else %
      \fi%
   \fi%
   \next%
}
\def\counthdROWS#1\into#2{%
\dvr{10}%
   \let\countREGISTER=#2%
   \countREGISTER=0%
\dvr{11}%
\dvr{13}%
   \expandafter\hdROWcount\the#1\endcount%
\dvr{12}%
}%
\def\hdROWcount{%
   \afterassignment\subhdROWcount\let\next= %
}%
\def\subhdROWcount{%
   \ifx\next\endcount %
      \let\next=\relax%
   \else%
      \ncase=0%
      \ifx\next\cr %
         \global\advance\countREGISTER by 1%
         \ncase=0%
      \fi%
      \ifx\next\endrow %
         \global\advance\countREGISTER by 1%
         \ncase=0%
      \fi%
      \ifx\next\crthick %
         \global\advance\countREGISTER by 1%
         \ncase=0%
      \fi%
      \ifx\next\crnorule %
         \global\advance\countREGISTER by 1%
         \ncase=0%
      \fi%
      \ifx\next\header %
         \ncase=1%
      \fi%
\relax%
      \ifcase\ncase %
         \let\next\hdROWcount%
      \or%
         \let\next\arghdROWskip%
      \else %
      \fi%
   \fi%
   \next%
}%
{\catcode`\|=13\letbartab
\gdef\countCOLS#1\into#2{%
   \let\countREGISTER=#2%
   \global\countREGISTER=0%
   \global\multispancount=0%
   \global\firstrowtrue
   \expandafter\COLcount\the#1\endcount%
   \global\advance\countREGISTER by 3%
   \global\advance\countREGISTER by -\multispancount
}%
\gdef\COLcount{%
   \afterassignment\subCOLcount\let\next= %
}%
{\catcode`\&=13%
\gdef\subCOLcount{%
   \ifx\next\endcount %
      \let\next=\relax%
   \else%
      \ncase=0%
      \iffirstrow
         \ifx\next& %
            \global\advance\countREGISTER by 2%
            \ncase=0%
         \fi%
         \ifx\next\span %
            \global\advance\countREGISTER by 1%
            \ncase=0%
         \fi%
         \ifx\next| %
            \global\advance\countREGISTER by 2%
            \ncase=0%
         \fi
         \ifx\next\|
            \global\advance\countREGISTER by 2%
            \ncase=0%
         \fi
         \ifx\next\multispan
            \ncase=1%
            \global\advance\multispancount by 1%
         \fi
         \ifx\next\header
            \ncase=2%
         \fi
         \ifx\next\cr       \global\firstrowfalse \fi
         \ifx\next\endrow   \global\firstrowfalse \fi
         \ifx\next\crthick  \global\firstrowfalse \fi
         \ifx\next\crnorule \global\firstrowfalse \fi
         \ifx\next\crnoruleneg \global\firstrowfalse \fi
         \ifx\next\crthickneg  \global\firstrowfalse \fi
         \ifx\next\crneg       \global\firstrowfalse \fi
      \fi
\relax
      \ifcase\ncase %
         \let\next\COLcount%
      \or %
         \let\next\spancount%
      \or %
         \let\next\argCOLskip%
      \else %
      \fi %
   \fi%
   \next%
}%
\gdef\argROWskip#1{%
   \let\next\ROWcount \next%
}
\gdef\arghdROWskip#1{%
   \let\next\ROWcount \next%
}
\gdef\argCOLskip#1{%
   \let\next\COLcount \next%
}
}
}
\def\spancount#1{
   \nspan=#1\multiply\nspan by 2\advance\nspan by -1%
   \global\advance \countREGISTER by \nspan
   \let\next\COLcount \next}%
\def\dvr#1{\relax}%
\def\header#1{%
\dvr{1}{\let\cr=\@mpersand%
\hdtks={#1}%
\counthdROWS\hdtks\into\hdrows%
\advance\hdrows by 1%
\ifnum\hdrows=0 \hdrows=1 \fi%
\dvr{5}\makehdPREAMBLE{\the\hdrows}%
\dvr{6}\getHDdimen{#1}%
{\parindent=0pt\hsize=\hdsize{\let\ifmath0%
\xdef\next{\valign{\headerpreamble #1\crnorm}}}\dvr{7}\next\dvr{8}%
}%
}\dvr{2}}
\def\makehdPREAMBLE#1{
\dvr{3}%
\hdrows=#1
{
\let\headerARGS=0%
\let\cr=\crnorm%
\edef\xtp{\vfil\hfil\hbox{\headerARGS}\hfil\vfil}%
\advance\hdrows by -1
\loop
\ifnum\hdrows>0%
\advance\hdrows by -1%
\edef\xtp{\xtp&\vfil\hfil\hbox{\headerARGS}\hfil\vfil}%
\repeat%
\xdef\headerpreamble{\xtp\crcr}%
}
\dvr{4}}
\def\getHDdimen#1{%
\hdsize=0pt%
\getsize#1\cr\end\cr%
}
\def\getsize#1\cr{%
\endsizefalse\savetks={#1}%
\expandafter\lookend\the\savetks\cr%
\relax \ifendsize \let\next\relax \else%
\setbox\hdbox=\hbox{#1}\newhdsize=1.0\wd\hdbox%
\ifdim\newhdsize>\hdsize \hdsize=\newhdsize \fi%
\let\next\getsize \fi%
\next%
}%
\def\lookend{\afterassignment\sublookend\let\looknext= }%
\def\sublookend{\relax%
\ifx\looknext\cr %
\let\looknext\relax \else %
   \relax
   \ifx\looknext\end \global\endsizetrue \fi%
   \let\looknext=\lookend%
    \fi \looknext%
}%
%
%
\def\tablelet#1{%
   \tableLETtokens=\expandafter{\the\tableLETtokens #1}%
}%
\catcode`\@=12